# A Bayesian approach to sharing information on sensitivity of a Multi-Cancer Early Detection test across and within tumour types and stages

Sofia Dias[1], Yiwen Liu[1], Stephen Palmer[2], Marta O Soares[2]

[1] Centre for Reviews and Dissemination, University of York, York, UK

[2] Centre for Health Economics, University of York, York, UK

Correspondence to: Sofia Dias, Centre for Reviews and Dissemination, University of York, York YO10 5DD, UK. Email: sofia.dias@york.ac.uk

**Competing interests**

The authors declare none.

**Data availability statement**

The data and code that support the findings of this study are openly available in GitHub at https://github.com/MCED-Galleri-HealthEconomicEval-Program/BayesianModelTestSens-1.

**Funding**

Financial support for this study was provided by NHS England (reference number C149303). The funding agreement ensured the authors' independence in designing the study, interpreting the data, writing, and publishing the report.

**Ethics**

This study used publicly available data only, therefore no ethics approval was required.

**Wordcount:** 5,203

**Abstract**

The Galleri® (GRAIL) multi-cancer early detection test measures circulating tumour DNA (ctDNA) to predict the presence of more than 50 different cancers, from a blood test. If sensitivity of the test to detect early-stage cancers is high, using it as part of a screening programme may lead to better cancer outcomes, but available evidence indicates there is heterogeneity in sensitivity between cancer types and stages.

We describe a framework for sharing evidence on test sensitivity between cancer types and/or stages, examining whether models with different sharing assumptions are supported by the evidence and considering how further data could be used to strengthen inference. Bayesian hierarchical models were fitted, and the impact of information sharing in increasing precision of the estimates of test sensitivity for different cancer types and stages was examined. Assumptions on sharing were informed by evidence from a review of the literature on the determinants of ctDNA shedding and its detection in a blood test.

Support was strongest for the assumption that sensitivity can be shared only across stage 4 for all cancer types. There was also support for the assumption that sensitivities can be shared across cancer types for each stage, if cancer types expected to have low sensitivity are excluded which increased precision of early-stage cancer sensitivity estimates and was considered the most appropriate model. High heterogeneity limited improvements in precision. For future research, elicitation of expert opinion could inform more realistic sharing assumptions.

**What is already known:** Bayesian information sharing models can be used to strengthen inferences by increasing precision of key estimates. However, sharing can be limited in the presence of unexplained heterogeneity.

**What is new:** We use a review of the literature to inform plausible sharing assumptions and implement them using three types of information sharing models applied to a dataset of the accuracy of a multi-cancer test across multiple cancer types, by stage of cancer at diagnosis.

**Potential impact for Research Synthesis Methods readers:** We present a novel framework to inform and implement modelling assumptions, present results and select the most appropriate model, applied to the synthesis of test accuracy data (sensitivity), which can be generalised to other cross-indication applications.

# 1 INTRODUCTION

When assessing health technologies, evidence on key parameters of interest is often sparse and heterogenous. However, relevant evidence from indirect sources, i.e. data that cannot be assumed to directly estimate a key parameter but can provide some relevant information on it, may be available. Bayesian information sharing models that allow the level of sharing to be moderated to reflect the support of the data and/or expert opinion for the assumptions, can be useful to synthesise evidence from indirect sources, whilst accounting for the heterogeneity across evidence sources.[1]

Multi-cancer early detection (MCED) tests are a novel technology that can detect potential signs of cancer before symptoms present based on a single sample of blood. Their use as part of a screening programme could lead to cancers being detected at an earlier stage (stage-shift), when treatment may be more effective and, perhaps, less costly compared to cancers detected at later stages, provided sensitivity of the test at early cancer stages is sufficiently high.

The Galleri® test (GRAIL, Menlo Park, California) is an MCED blood test which uses genetic sequencing to detect circulating tumour DNA (ctDNA) that signals the presence of cancer. The test looks for a signal shared by more than 50 different types of cancer and is currently being trialled in the UK for repeat screening in the general population.[2]

A systematic review of accuracy evidence for MCED tests[3] identified, quality appraised and summarised evidence on 13 different technologies. Despite identifying limited evidence test sensitivity was found to vary across different technologies, cancer types and stages.[3] Three of the studies identified evaluated the performance of the Galleri MCED test; these differed in design and included populations and none included high-quality evidence in a screening population.[3] Two were prospective cohort studies which recruited 6621 adults aged 50 and over with or without elevated risk of cancer in the US (of which 120 were found to have cancer, PATHFINDER),[4] and 5461 adults aged 18 and over who were referred for urgent investigation of cancer symptoms in the UK (of which 368 were found to have cancer, SYMPLIFY).[5] CCGA sub-study 3 (CCGA3)[6] was a case-control study of adults aged 20 and over with (2823 individuals) and without (1254 individuals) cancer in North America. Overall, no high-quality evidence on the sensitivity of the Galleri test in a screening population was available and diagnostic accuracy varied substantially, with sensitivity pooled across all cancer types and stages varying from 20.8% in PATHFINDER, 51.5% in CCGA3, to 66.3% in SYMPLIFY, but generally high specificity (98.4% to 99.5%).[3] Sensitivity was also generally lower for detecting earlier stage cancers compared with later stage cancers.[3]

Accuracy studies showed variation in test sensitivity across cancer types; this is expected due to known differences in ctDNA expression.[7-9] However, for a large proportion of the cancer types detected by the Galleri test, the numbers of observed patients in existing studies are small, leading to significant uncertainty in test sensitivity.[3] Where the Galleri test is expected to have similar sensitivity between two or more tumour types, or across different stages, the data collected on one cancer type at a particular stage, could be used alongside the data collected on another to share information and strengthen inferences.

Bayesian information sharing models can be used to synthesise evidence across tumour types and/or stages of disease.[1] These models do not need to assume that evidence from different sources is estimating the exact same parameter, but can impose different levels of sharing information based on different assumptions about the relationships across the evidence sources.[1, 10] Structural and exchangeability-based relationships can be used and combined with prior information, but unexplained heterogeneity can limit precision gains.

We will explore the use of Bayesian information sharing models applied to data on sensitivity of the Galleri MCED test, to examine support for sharing evidence across cancer types and cancer stages, and improvements in precision. In Section 2 we describe the available data on test sensitivity for the Galleri test and consider existing evidence on sources of heterogeneity in ctDNA-based MCED test performance to support sharing of information across cancer types and stages. In Section 3 we introduce the models used to statistically examine the levels of heterogeneity in test sensitivity for different tumour types and stages. In Section 4 we identify the sharing assumptions that retain plausibility given the existing data and explore the impact of different levels of information sharing on increasing precision of the estimates. We conclude with a discussion of the results and ongoing research.

## 2  AVAILABLE EVIDENCE

### 2.1  *Sensitivity of the Galleri test*

To minimise heterogeneity due to different study designs, and because cancer- and stage-specific data from other studies is limited, only data from CCGA3 will be used for the exploration of sharing models, as this study detected the most cancers at various stages. Data on sensitivity by cancer type and stage for CCGA3[6] are given in Table 1. Table S1 gives the 95% confidence intervals (CIs). Unstageable cancers (lymphoid leukaemia and myeloid neoplasm) and cancers where information on stage is missing were removed from the synthesis dataset. Figure 1 and Figure 2 show the observed sensitivities (dot) and 95% CIs (solid lines) for each cancer type and stage. Observed sensitivities for non-specific cancer types (labelled "other", "unknown" or "multiple") are presented in Figure S1. There is clear heterogeneity between cancer types, for example stage 1 sensitivity varies between 0% (e.g. thyroid) and 100% (e.g. liver/bile duct). In addition, it is noted that even though sensitivity is expected to increase with stage (a particular tumour that progresses in stage over time sheds increasing ctDNA),[11, 12] because of uncertainty, this is not always observed in the data.

Breast, prostate, lung and colon/rectum represent a large proportion (61.2%) of the included and detected cancers at all stages. These cancers will contribute most to the estimation of sensitivity in models that allow information sharing. The included sample for breast and prostate cancer had mainly early-stage disease whereas the samples for lung and colon/rectum included mostly advanced cancer and the patterns of stage-specific sensitivity across these cancer types is heterogeneous (Table 1). Other cancer types had fewer cancers detected at each stage and may benefit most from models that allow information sharing (i.e. sensitivity estimates may gain precision), although conflict between model predictions and the evidence needs to be assessed. Thyroid, melanoma and urothelial tract had no cancers detected at stages 1-3 (Table 1) so these cancer types will not contribute much to the estimation of stage-specific sensitivities but it is expected these estimates will gain precision when there is information sharing, although conflict between model estimates and the evidence is unlikely to be detected.

## *2.2 Clinical determinants of heterogeneity*

Characteristics that could determine the potential homogeneity or heterogeneity of the sensitivity of the Galleri test across different cancer types and stages were identified from targeted searches of the literature. This (indirect) evidence was then used to inform plausible information sharing assumptions to be implemented in the information sharing models. A description of the review methods is provided in Supplement S2.

There is consistent evidence that ctDNA shedding is higher in advanced stages of cancer with higher tumour burden compared to localised cancer, regardless of the type of cancer.[7, 11-17] The size of the tumour (which typically reflects a more advanced stage) has also been associated with the detectability of ctDNA in blood samples.[7, 18] Thus, tumour ctDNA shedding may be the main driver of differences in test performance.[8, 9] There is also evidence to suggest that ctDNA levels measured in blood may be lower in cancers of the central nervous system (e.g., glioma, medulloblastoma) due to the blood-brain barrier.[11, 12, 19, 20] Lower levels of ctDNA have also been found in renal cell, bladder, and kidney cancer,[13, 19, 21] possibly due to the clearance of ctDNA via urine.[22] Thyroid cancer has also been shown to have a lower level of ctDNA.[11-13, 19]

Other tumour characteristics, clinical factors and population demographic factors have also been associated with levels of ctDNA, although none of these factors are available in the Galleri dataset so are not explored further. Further details are available in Supplement S2.

# 3 METHODS

Bayesian hierarchical models with different structural assumptions for sharing information across stages and cancer types are proposed. Data from each study $i$ are the number of cancers of type $j$ at stage $k$ detected, $s_{ijk}$ (true positives), out of the total number of individuals with that cancer type and stage, $S_{ijk}$, assumed to follow a Binomial distribution with probability (test sensitivity) $p_{ijk}$, modelled on the log-odds scale

$$s_{ijk} \sim \text{Binomial}(p_{ijk}, S_{ijk})$$
$$\text{logit}(p_{ijk}) = \delta_{ijk} \quad (1)$$

for $i=1,\ldots, N$ (the total number of studies to be pooled), $j=1,\ldots, J_i$ (the total number of cancer types observed in study $i$) and $k=1,\ldots,4$ (the number of cancer stages).

The log-odds of the probabilities of detecting a true cancer of a given type and stage in each study, $\delta_{ijk}$, can be assumed to be common across studies (common effect model)

$$\delta_{ijk} = \mu_{jk} \quad (2)$$

to estimate a single pooled log-odds of detection for each cancer type and stage across all studies, or they can be assumed to come from a common distribution, for example Normal with mean $\mu_{jk}$ and variance $\sigma^2$, which describes the between-study heterogeneity (random effects model):

$$\delta_{ijk} \sim \text{Normal}(\mu_{jk}, \sigma^2) \quad (3)$$

Assumptions of information sharing across cancer types, $j$, and stages, $k$, are imposed on $\mu_{jk}$.

Although the model is set up to synthesise evidence across more than one study, in this paper N=1 (only CCGA3 data are used), and only common effect models are considered.

## 3.1 Base model

Cancers at more advanced stages are expected to shed more ctDNA and are more likely to be detected by the Galleri test. Our base model will constrain the sensitivities across cancer types and stages to be independent but monotonically increasing (or equal) with increasing stage, whilst fully accounting for the uncertainty in the data:[23, 24] for all cancer types $j$, we ensure that $\mu_{j1} \leq \mu_{j2} \leq \mu_{j3} \leq \mu_{j4}$. For further details on implementation of the constraints see Supplement S3. Independent, non-informative prior distributions are specified for all cancer types $j$ at stage $k$

$$\mu_{jk} \sim \text{Normal}(0, 100^2) \tag{4}$$

Including constraints implies some information sharing across adjacent cancer stages within-cancer type, which will lead to an increase in precision of estimates across stages for each cancer type compared to a model with no constraints.

Because this model reflects typical assumptions in evidence synthesis in this area, and mimics previous analyses of these same data,[25, 26] it will serve as a benchmark against which to measure precision increases and changes to point estimates for models that share information across cancer types and stages more explicitly.

### *3.2 Information sharing models*

The base model is extended to allow different levels of sharing of information between cancer types, based on 3 types of Bayesian models[27-29] which implement alternative sharing assumptions in addition to the monotonically increasing constraints.

Exchangeability models allow the probabilities of detecting cancer at particular stages to be assumed to be similar (exchangeable) across some cancer types whilst allowing for heterogeneity between the different cancer types. The log-odds of the sensitivities for each stage come from a common distribution, with mean $m_k$ and variance $\tau_k^2$ (measuring heterogeneity across different cancer types), i.e. they are exchangeable across cancer types, but can be different across stages:

$$\mu_{jk} \sim \text{Normal}(m_k, \tau_k^2) \tag{5}$$

If exchangeability is only appropriate across certain cancer types, these can be specified in a set $V$ and equation (5) applied only to $j$ in set $V$, with prior distributions in equation (4) applied to cancer types $j$ not it set $V$ (i.e. no sharing). The latter cancer types will only have sensitivities constrained to increase with stage, but no other borrowing of information is imposed. Non-informative and minimally informative prior distributions are given for each $m$ and $\tau$, respectively:

$$\begin{aligned} &m_k \sim \text{Normal}(0, 100^2) \,, \; \tau_k \sim \text{Uniform}(0, 5) \\ &k = 1, \ldots, 4 \end{aligned} \tag{6}$$

This model can be adapted to allow for common means or heterogeneity across some or all stages by specifying equality between some (or all) $m_k$ and/or $\tau_k$ and adjusting the prior distributions accordingly.

Mixture models explicitly consider how much each individual cancer type contributes to the sharing element, using a (mixture) probability parameter which moderates how much, for a given stage, each particular cancer type contributes to the sharing element.[30, 31] A low probability is estimated when the evidence strongly indicates that a cancer type differs from others which are more similar between themselves (i.e. for which there is no strong evidence of difference). A high probability indicates no evidence of a difference between a particular cancer type and the others, so information can be shared. Thus, mixture models can prevent too much sharing from "extreme" cancer types and stages (i.e. where sensitivities are very different), reducing the contribution of extreme indications to the heterogeneity, and strengthening sharing within the cancer types that are more similar.[31]

Variable $X$ determines whether the sensitivity for a particular cancer type and stage is exchangeable with the rest according to a probability $\pi$ which is given a Beta prior distribution:

$$X_{jk} \sim \text{Bernoulli}(\pi_{jk})$$
$$\pi_{jk} \sim \text{Beta}(a_{jk}, b_{jk}) \tag{7}$$

Prior distributions for $\pi_{jk}$ can be set as non-informative ($a=b=1$), or informed by elicitation of expert opinion or other external evidence.

The log-odds of the sensitivities for each stage are assumed to follow equation (5) when $X_{jk}=1$, or equation (4) when $X_{jk}=0$. The posterior mean of $X$ will indicate the probability that the stage-specific sensitivity for cancer type $j$, stage $k$ is exchangeable with the stage-specific sensitivities of the other cancer types, as determined by the data. The mixture models sharing assumption can be applied to one or more subsets of cancer types for which sharing is plausible, leaving others to be estimated independently.

Two mixture models will be considered: one where the mixture probabilities depend on cancer type and stage (equation (7)) and another where there is a single mixture probability across all stages of the same cancer type (see Supplement S3 for details). Prior distributions for the pooled means and heterogeneity parameters are given in equation (6).

Class models explicitly include external evidence suggesting that some cancer types are more similar to certain others in terms of expected sensitivity by stage, by defining several groups (classes) of cancers within which sharing stage-specific sensitivity is appropriate, but where sharing across classes is not allowed.[24, 32, 33] The log-odds of the sensitivities for each stage within pre-defined cancer groups (classes) indexed by a vector $D$, are assumed to come from a common distribution. Equation (5) is replaced by:

$$\mu_{jk} \sim \text{Normal}\left(m_{D_j,k}, \tau^2_{D_j,k}\right) \tag{8}$$

Non-informative and minimally informative prior distributions are given for each $m$ and $\tau$

$$\begin{aligned} m_{ck} &\sim \text{Normal}(0, 100^2), \quad c = 1,\ldots, nClass;\ k = 1,\ldots, 4 \\ \tau_{ck} &\sim \text{Uniform}(0, 5) \end{aligned} \tag{9}$$

where *nClass* defines the number of pre-specified cancer groups (classes).

### 3.3  Model specification

As ctDNA independently predicts test performance for different stages (Section 2.2), assumptions on similarities, or differences, between sensitivities for different cancer types and stages will reflect what is known about the heterogeneity in ctDNA shedding of each cancer type at each stage. Cancer types described as "Other", "Unknown" or "Multiple" are not included in any sharing assumptions, since they comprise mixtures of different cancer types for which it would be difficult to determine appropriate sharing assumptions and interpret results. Results for these cancer types are only presented for the base model (see Figure S1).

It is expected that the Galleri test will have different sensitivity depending on the stage the cancer is presenting at, but it may be reasonable to assume that sensitivity for each stage is similar across all cancer types. These assumptions are implemented in Models 1A and 1B.

Evidence also suggests that the Galleri test will have higher sensitivity for stage 4 cancers than for cancers at stages 1-3. These assumptions are implemented in Model 2.

Bladder, kidney and thyroid cancer may have a lower probability of being detected by the Galleri test at all stages so it may be reasonable to assume that information from these cancers should not be shared with the other cancer types. These assumptions are implemented in Model 3.

Breast, lung and colorectal cancer may have higher sensitivity at stage 4 than expected for most other cancer types. To prevent "high stage 4 sensitivity" cancer types from artificially inflating stage 4 sensitivity for other cancer types, Model 4 shares sensitivity of stage 4 across all cancers except breast, lung and colon/rectum.

Model 5 is a mixture model where mixture probabilities depend on cancer type and stage, in Model 6 a single mixture probability across all stages of the same cancer type is assumed. Both models use non-informative (Beta(1,1)) prior distributions for the mixture probabilities.

Model 7 is a class model with groupings based on results from the literature review supported by the results of the mixture models.

Table 2 gives details of the implemented models.

Cancer groups based on tumours that share similar dwell times,[25] expected late-stage incidence reduction (stage-shift)[26] or 5-year predicted survival rates,[34] have been suggested in previous studies. These characteristics may be related to levels of ctDNA shedding, although this was not the explicit motivation for these groupings. Additional cancer groups were formed by using the data in Table 1 to identify clusters of cancer based on K-means clustering (for details see Supplement S4). In sensitivity analyses class models using these alternative groupings were fitted. Table S2 provides details of all the cancer groups considered.

### 3.4   Model selection and fit

Statistical models with different information sharing assumptions are compared based on measures of model fit,[35, 36] the magnitude of estimated heterogeneity across stages or cancer types and clinical plausibility of model predictions given by the (shrunken) estimates for each cancer type and stage. Models with adequate fit (based on the residual deviance) and predictions (based on the shrunken estimates), were compared by looking at differences in Deviance Information Criteria (DIC).[35, 36] Models with similar residual deviance but lower DIC were preferred. Models were estimated using Markov chain Monte Carlo (MCMC) implemented in JAGS. For further details on model implementation and selection see Supplement S3.

## 4   RESULTS

All models in Table 2 were fitted to the data in Table 1. Model fit statistics are presented in Table 3. Individual data points' deviance contributions are available online at https://github.com/MCED-Galleri-HealthEconomicEval-Program/BayesianModelTestSens-1.

Figure 1 and Figure 2 present the model predictions for each cancer type and stage (shrunken estimates) as density strips (plotted using R package 'denstrip', version 1.5.4) with vertical ticks denoting the lower bound of the 95% credible interval (2.5% quantile), the median and the upper bound of the 95% credible interval (97.5% quantile), respectively, for the base model and the four best-fitting models (Models 2 to 5). See Figures S2 and S3 for results of the other fitted models.

The base model fits the data well (Table 3) and improves precision of the sensitivity estimates (Figure 1, Figure 2 and Table S5), particularly when the total number of cancers detected was low and estimates across consecutive stages are close (e.g. breast stage 4, colon/rectum stages 3 and 4, sarcoma all stages, Figure 1). Where the observed sensitivities were not increasing with stage (e.g.

bladder stages 1-2, kidney stages 2-3, sarcoma stages 2-3, Figure 1) the base model generates increasing estimates without introducing conflict with the observed data.

In Models 1A and 1B pooled sensitivities across stages were similar (Supplementary Figures S2, S3 and Table S6) but heterogeneity was high (Table S6). However, for both models, fit was poor with residual deviance over 10 points higher than the number of data points (Table 3). These models allow for most sharing but the poor fit and substantial difference in model estimates for some cancer types compared to the base model, suggest that more refined sharing assumptions are needed. As Model 1A and 1B have similar fit and consistent heterogeneity estimates, suggesting that a common heterogeneity parameter is a reasonable assumption. Therefore, all other models will assume that within cancer type heterogeneity is common across stages (Model 1B).

Model 2 fitted the data well, with residual deviance similar to the base model (Table 3). This indicates that the data supports sharing of information for stage 4 sensitivity across cancer types. A high pooled sensitivity is estimated for stage 4 across all cancer types, although heterogeneity was high (Table S6). The high estimated heterogeneity moderates sharing so stage 4 estimates are not changed much compared to the base model for cancers with substantial evidence on stage-specific sensitivity (e.g. colon/rectum, head and neck, lung, lymphoma, pancreas, Figure 1). Cancer types with few observations at stage 4 have a greater shift in the estimated sensitivity compared to the base model, although the high level of between-cancer heterogeneity does not allow for much gain in precision (e.g. thyroid cancer, Figure 2).

Model 3 has acceptable model fit (Table 3) although within-stage heterogeneity across the different cancer types is high, so precision gains compared to the base model are limited (Table S6). The improvement in fit for Model 3 compared to Models 1A and 1B is substantial (difference in residual deviance > 8 points, difference in DIC > 15 points, Table 3), suggesting that sharing information across stages may be more reasonable when "low sensitivity" cancer types are excluded.

Model 4 fitted the data well (Table 3) although the improvement in fit compared to Model 2 was modest and model predictions for each cancer type and stage were similar (Figure 1, Figure 2 and Table S6). Higher pooled stage 4 sensitivity and heterogeneity than Model 2 (Table S6) contradict expectations that pooled stage 4 sensitivity might have been inflated by inclusion of these "high sensitivity" cancers in Model 2. This provides support for sharing stage 4 sensitivity across *all* cancer types (Model 2). Note that there are several cancer types apart from the suggested "high sensitivity" cancers, where 100% of stage 4 cancers were detected (Table 1).

Estimated mixture probabilities for Models 5 and 6 are given in Supplementary Table S7. Model 5 fitted the data well, but Model 6 had high residual deviance compared to the number of data points (Table 3) and convergence of key parameters was poor. Model 5 and 6 pooled sensitivity estimates are presented in Table S6.

Some "extreme" cancer type/stage combinations (probability of exchangeability < 0.5) were identified in Model 5 (Table S7). The estimated sensitivities for cancer types/stages with low probability of exchangeability are very close to those of the base model, as sharing is limited (e.g. prostate cancer stage 3). For other cancer types and stages, estimates of sensitivity tend to be closer to those of Model 3 where information is shared across cancer types for each stage. However, some differences are notable: the estimated sensitivities for bladder and kidney cancer from Model 5 are very different from those of the base model and Model 3 because in Model 5 these cancer types are allowed to share information with other cancer types (for each stage) in the latter but nor the former (Figure 1 and Table S6).

The low mixture probability for melanoma, gallbladder and urothelial tract in model 5 may be due to the small number of early-stage cancers included in the study so the finding that these cancer types are extreme may be unreliable. On the other hand, prostate cancer has low sensitivity for stages 2 and 3, compared to most other cancers, and has many observed cancers at these stages, which may explain the lower estimated mixture probabilities. All the cancer types with low mixture probabilities for stage 4 (and cervical cancer at stage 3) in Model 5 had 100% sensitivity based on a small number of cases. These findings support the assumptions that the "low sensitivity" cancer types may be different from the others (Model 3), although bladder and kidney cancer were not identified as different in the mixture model (Model 5) due to the small number of observations and wide uncertainty in the estimates. The cancer types suggested to have "high stage 4 sensitivity" were not identified as different by the mixture model, reinforcing the previous conclusion that this assumption is not supported by the data.

In Model 7 within-class/group heterogeneity is lower than in other sharing models (Table S8), but the fit of this model was poor (Table 3) suggesting that sharing of information within these pre-defined groups may not be appropriate. Model estimates for cancer types in Group 1 and some of the cancers in Group 2 differ from the base model and are often more precise (Figures S2 and S3), which accounts for some of the poor model fit.

## 4.1 Model selection

Although the base model had the best fit overall, models that include plausible sharing assumptions to improve precision whilst still fitting the data well are preferable. The fit of Model 2 was comparable to Model 4 so relaxing the assumption of sharing over "high sensitivity" cancers was not considered necessary, therefore Model 2 is preferred over Model 4. However, Model 2 does not allow sharing over early-stages of cancer, where precision gains are likely to have greater value. Model 3 has more useful sharing assumptions and is selected as an alternative plausible model, even though the fit is slightly worse.

## *4.2 Sensitivity analyses*

All models using the alternative cancer groups in Table S2 improved the fit over the Model 7, although no model had a good fit to the data (Table S9). Estimated sensitivities and heterogeneity are shown in Table S10. Overall these analyses do not impact our choice of preferred models.

# 5 DISCUSSION

Using the largest study on sensitivity of the Galleri test, CCGA3, we conducted an exploratory analysis of determinants of heterogeneity across tumour types and stages. We found substantial heterogeneity of test sensitivity across cancer types and stages, particularly for detection of early-stage cancers, suggesting that there are some cancer types that the Galleri test is more likely to find early than others. However, more data are needed to conclusively identify which cancer types are more likely to be detected early by the Galleri test. Until more is known, sharing of evidence using flexible models will be required to accommodate uncertainty. Our work shows that models share information only weakly but still provide some increases in precision.

Mean sensitivities estimated by our base model agree with those used in a previous modelling study (Table S5) although uncertainty measures were not presented.[26] In fitting information sharing models, we found support was strongest for the assumption that sensitivity of the Galleri test can be shared across stage 4 for all cancer types. However, movement from detection of cancers at stage 4 to detection at earlier stages (stage-shift), is the most important determinant of value of this type of test when used for screening. There was significant heterogeneity within the other stages, and while our explorations (based on current evidence) explained some of the heterogeneity, remaining unexplained heterogeneity was still high.

Our explorations were based on current evidence on the determinants of ctDNA shedding and detection in a blood test, which is mostly focussed on a few of the most common cancers and in comparing late-stage vs early-stage rather than providing stage-specific evidence. As research on this topic advances and our understanding of ctDNA in the blood and its relationship to MCED tests increases, more refined assumptions can be proposed in the synthesis of these data. We have demonstrated that there can be precision gains in the estimates of test sensitivity by stage and tumour type by constraining values to increase with stage for each tumour type. After constraining estimates, sharing stage-specific estimates across tumour types using flexible sharing models is plausible but, for this to lead to further precision gains, expert judgement is required to support the identification of cancer types across which sharing is most clinically plausible. Further understanding of the heterogeneity would allow models with stronger sharing assumptions to be implemented, and evidence to be used more efficiently with relevant impacts for policy. In the absence of empirical evidence, further research could seek expert opinion to support more complex, but also more realistic, sharing assumptions. This could include eliciting support for groupings of cancer types, or eliciting prior distributions for sharing probabilities for the mixture models. Expert opinion should also confirm the validity of the assumption that sensitivity increases with increasing stage at the aggregate level, which has support from CCGA3 data.

Evidence on sensitivity of the Galleri test was available from two additional studies: SYMPLIFY and PATHFINDER. However, current evidence was insufficient to examine differences in sensitivity across CCGA3, SYMPLIFY and PATHFINDER studies. Given the large variability in study design and the much larger number of cancers detected in CCGA3, there would be little additional information gained from an analysis that pooled data from the three studies. Data sparseness would also prevent the exploration of heterogeneity of sensitivity estimates across studies. However, the models proposed here can be extended to include multiple studies and this can be explored as more evidence becomes available.

The ongoing NHS-Galleri randomised controlled trial[2] will provide information in a relevant UK screening population. However, because participants are asymptomatic and the trial is powered on overall stage-shift across all cancer types, for some cancer types, few cancers may be detected, particularly at early stages, so the NHS-Galleri trial may not be able to robustly inform stage-shift for all the different cancer types. Whilst we expected overall sensitivity to differ with different study designs, conditional on stage and tumour type, it is possible that sensitivities may be similar across different study types and that conditional sensitivities may be generalisable across different study designs and populations. This could be explored by incorporating tumour and stage-specific data from NHS-Galleri with the existing Galleri evidence or use it for bias-adjustment.

An economic model is being developed to provide an initial estimate of the cost-effectiveness of a screening programme using the Galleri test in the UK NHS, grounded on the primary results of the NHS-Galleri trial. It is important that the evidence in the economic model that determines the attribution of the overall stage shift results from the NHS-Galleri trial across cancer types is robust. A key parameter for such attribution is the sensitivity of the Galleri test in detecting each different cancer type at different stages. This parameter, together with the speed of pre-clinical progression, will determine the likelihood and ability of the Galleri test to impact on stage-shift for each cancer type. Our sharing models can strengthen the evidence on the sensitivity of the Galleri test across tumour types and stages, whilst quantifying heterogeneity between cancer types. A high specificity is equally important to minimise the number of false positive results which can incur costs, stress and time burden. Specificity of the Galleri test was found to be high although there was insufficient information on the potential cost and healthcare implications of false positives.[3]

We considered models that imposed sharing assumptions on the stage-specific sensitivities (log-odds scale) for each cancer type. Alternative models could have been used, for example, we could have expressed test sensitivity at a particular stage as an increase in sensitivity from the previous stage and shared the increments (rather than absolute values) across tumour types. However, because the increments are on a log-odds scale but must also be positive (to impose the constraint of increasing sensitivity with stage) implementing sharing assumptions on the increments and interpreting the results would be challenging.

Although improving the precision of sensitivity estimates for the Galleri MCED test was the motivation for this work, the modelling approach proposed here can also be used to explore heterogeneity and appropriate sharing assumptions for cross-indication diagnostic tests, to better inform decision making.

**CRediT**

Sofia Dias: Conceptualization, Data curation, Formal analysis, Funding acquisition, Methodology, Software, Supervision, Validation, Visualization, Writing – original draft, Writing – review & editing.

Yiwen Liu: Data curation, Formal analysis, Methodology, Validation, Writing – review & editing.

Stephen Palmer: Funding acquisition, Methodology, Project administration, Supervision, Writing – review & editing.

Marta O Soares: Conceptualization, Data curation, Formal analysis, Funding acquisition, Methodology, Project administration, Validation, Writing – review & editing.

# TABLES AND FIGURES

**Table 1. CCGA3 sensitivity data by cancer type and stage**

| Code | Cancer type | Sensitivity by cancer type and stage, CCGA3; s/S (%) | | | | | |
|---|---|---|---|---|---|---|---|
| | | Overall | Stage I | Stage II | Stage III | Stage IV | Missing |
| 1 | Bladder | 8/23 (34.8) | 2/6 (33.3) | 1/11 (9.1) | 3/4 (75.0) | 2/2 (100.0) | 0/0 (NA) |
| 2 | Breast | 160/524 (30.5) | 7/265 (2.6) | 86/181 (47.5) | 47/55 (85.5) | 20/22 (90.9) | 0/1 (0.0) |
| 3 | Colon/Rectum | 169/206 (82.0) | 13/30 (43.3) | 34/40 (85.0) | 58/66 (87.9) | 61/64 (95.3) | 3/6 (50.0) |
| 4 | Head and neck | 90/105 (85.7) | 12/19 (63.2) | 14/17 (82.4) | 16/19 (84.2) | 48/50 (96.0) | 0/0 (NA) |
| 5 | Kidney | 18/99 (18.2) | 3/61 (4.9) | 2/9 (22.2) | 1/7 (14.3) | 12/22 (54.5) | 0/0 (NA) |
| 6 | Liver/bile duct | 43/46 (93.5) | 6/6 (100.0) | 7/10 (70.0) | 9/9 (100.0) | 20/20 (100.0) | 1/1 (100.0) |
| 7 | Lung | 302/404 (74.8) | 21/96 (21.9) | 35/44 (79.5) | 107/118 (90.7) | 138/145 (95.2) | 1/1 (100.0) |
| 8 | Lymphoma | 98/174 (56.3) | 9/33 (27.3) | 28/48 (58.3) | 33/46 (71.7) | 28/46 (60.9) | 0/1 (0.0) |
| 9 | Ovary | 54/65 (83.1) | 5/10 (50.0) | 4/5 (80.0) | 27/31 (87.1) | 18/19 (94.7) | 0/0 (NA) |
| 10 | Pancreas | 113/135 (83.7) | 13/21 (61.9) | 12/20 (60.0) | 18/21 (85.7) | 70/73 (95.9) | 0/0 (NA) |
| 11 | Prostate | 47/420 (11.2) | 3/95 (3.2) | 12/243 (4.9) | 7/50 (14.0) | 25/30 (83.3) | 0/2 (0.0) |
| 12 | Sarcoma | 18/30 (60.0) | 4/10 (40.0) | 2/2 (100.0) | 5/10 (50.0) | 6/7 (85.7) | 1/1 (100.0) |
| 13 | Thyroid | 0/14 (0.0) | 0/11 (0.0) | 0/1 (0.0) | 0/1 (0.0) | 0/1 (0.0) | 0/0 (NA) |
| 14 | Uterus | 44/157 (28.0) | 20/120 (16.7) | 3/10 (30.0) | 17/23 (73.9) | 4/4 (100.0) | 0/0 (NA) |
| 15 | Lymphoid leukaemia[a] | 21/51 (41.2) | | | | | |
| 16 | Melanoma | 6/13 (46.2) | 0/2 (0.0) | 0/2 (0.0) | 0/3 (0.0) | 6/6 (100.0) | 0/0 (NA) |
| 17 | Plasma cell neoplasm | 34/47 (72.3) | 11/17 (64.7) | 14/16 (87.5) | 9/14 (64.3) | 0/0 (NA) | 0/0 (NA) |
| 18 | Anus | 18/22 (81.8) | 1/4 (25.0) | 3/4 (75.0) | 13/13 (100.0) | 1/1 (100.0) | 0/0 (NA) |
| 19 | Cervix | 20/25 (80.0) | 7/12 (58.3) | 5/5 (100.0) | 7/7 (100.0) | 1/1 (100.0) | 0/0 (NA) |
| 20 | Gallbladder | 12/17 (70.6) | 0/2 (0.0) | 1/3 (33.3) | 3/4 (75.0) | 8/8 (100.0) | 0/0 (NA) |
| 21 | Urothelial tract | 8/10 (80.0) | 0/2 (0.0) | 0/0 (NA) | 0/0 (NA) | 8/8 (100.0) | 0/0 (NA) |
| 22 | Myeloid neoplasm[a] | 2/10 (20.0) | | | | | |
| 23 | Oesophagus | 85/100 (85.0) | 1/8 (12.5) | 11/17 (64.7) | 32/34 (94.1) | 40/40 (100.0) | 1/1 (100.0) |
| 24 | Stomach | 20/30 (66.7) | 1/6 (16.7) | 3/6 (50.0) | 4/5 (80.0) | 12/12 (100.0) | 0/1 (0.0) |
| 25 | Other[b] | 30/59 (50.9) | 2/11 (18.2) | 3/3 (100.0) | 13/18 (72.2) | 11/18 (61.1) | 1/3 (33.3) |
| 26 | Multiple primaries[c] | 16/19 (84.2) | 2/2 (100.0) | 3/5 (60.0) | 6/6 (100.0) | 5/6 (83.3) | 0/0 (NA) |
| 27 | Unknown primary | 17/18 (94.4) | 0/0 (NA) | 1/1 (100.0) | 1/2 (50.0) | 13/13 (100.0) | 2/2 (100.0) |
| | **Total** | **1453/2823 (51.5)** | **143/849 (16.8)** | **284/703 (40.4)** | **436/566 (77.0)** | **557/618 (90.1)** | **10/20 (50.0)** |

[a] Not expected to be staged.
[b] Includes: adrenal (n=1); ampulla of vater (n=1); brain (n=6); choriocarcinoma (n=1); mesothelioma (n=7); non-melanoma non-BCC/SCC skin cancer (n=2); other/unspecified (n=10); penis (n=1); small intestine (n=13); testis (n=6); thymus (n=2); vagina (n=2); vulva (n=7). Also includes 6 which were not expected to be staged and therefore excluded from data by stage, but still counted in the overall column.
[c] Multiple primaries: If participant had more than 1 cancer at enrolment. If they had multiple primaries and unknown primary, they were counted as multiple primaries to avoid double counting. Highest clinical stage was selected.

**Table 2 Description of modelling assumptions**

|  | **Assumptions on sensitivity** [a] | **Assumption on heterogeneity** |
|---|---|---|
| **Model 1** | Exchangeability model: Sensitivity is shared across all cancer types for each stage | A) common within stage (across cancer type) heterogeneity<br>B) separate heterogeneity across cancer types for each stage |
| **Model 2** | Exchangeability model: Only sensitivity for stage 4 is shared across all cancer types | common within stage (across cancer type) heterogeneity |
| **Model 3** | Exchangeability model: Sensitivity for each stage shared across cancer types, except for "low sensitivity" cancers (bladder, kidney and thyroid) | common within stage (across cancer type) heterogeneity |
| **Model 4** | Exchangeability model: Sensitivity for stage 4 is shared across cancer types, except "high stage 4 sensitivity" cancers (breast, lung and colon/rectum) | common within stage (across cancer type) heterogeneity |
| **Model 5** | Mixture model: mixture probabilities depend on cancer type and stage | A) sharing component allows for between cancer type heterogeneity<br>B) sharing component assumes common effects |
| **Model 6** | Mixture model: single mixture probability across all stages of the same cancer type | sharing component allows for between cancer type heterogeneity |
| **Model 7** | Class model: group 1 = bladder, kidney, prostate, thyroid, melanoma (low sensitivity at all stages); group 2 = remaining cancer types. | Within class/group heterogeneity common across stages. |

[a] All models assume sensitivity is increasing with stage for each cancer type

**Table 3 Model fit statistics**

|  | **Residual deviance**[1] | **Deviance** | **pD** | **DIC** |
|---|---|---|---|---|
| **Base model** | 92.7 | 287.0 | 74.3 | 361.4 |
| **Model 1A** | 117.8 | 312.1 | 88.3 | 400.3 |
| **Model 1B** | 118.8 | 313.1 | 88.3 | 401.4 |
| **Model 2** | 98.6 | 292.9 | 76.8 | 369.7 |
| **Model 3** | 109.7 | 304.0 | 80.1 | 384.0 |
| **Model 4** | 97.1 | 291.4 | 77.0 | 368.4 |
| **Model 5** | 103.6 | 297.9 | 90.9 | 388.8 |
| **Model 6** | 111.5 | 305.8 | 86.5 | 392.3 |
| **Model 7** | 122.4 | 316.7 | 85.1 | 401.8 |

[1] compare to 100 data points

**Figure 1 Data and shrunken estimates of sensitivity for the base model and models 2 to 5 (Part 1)**

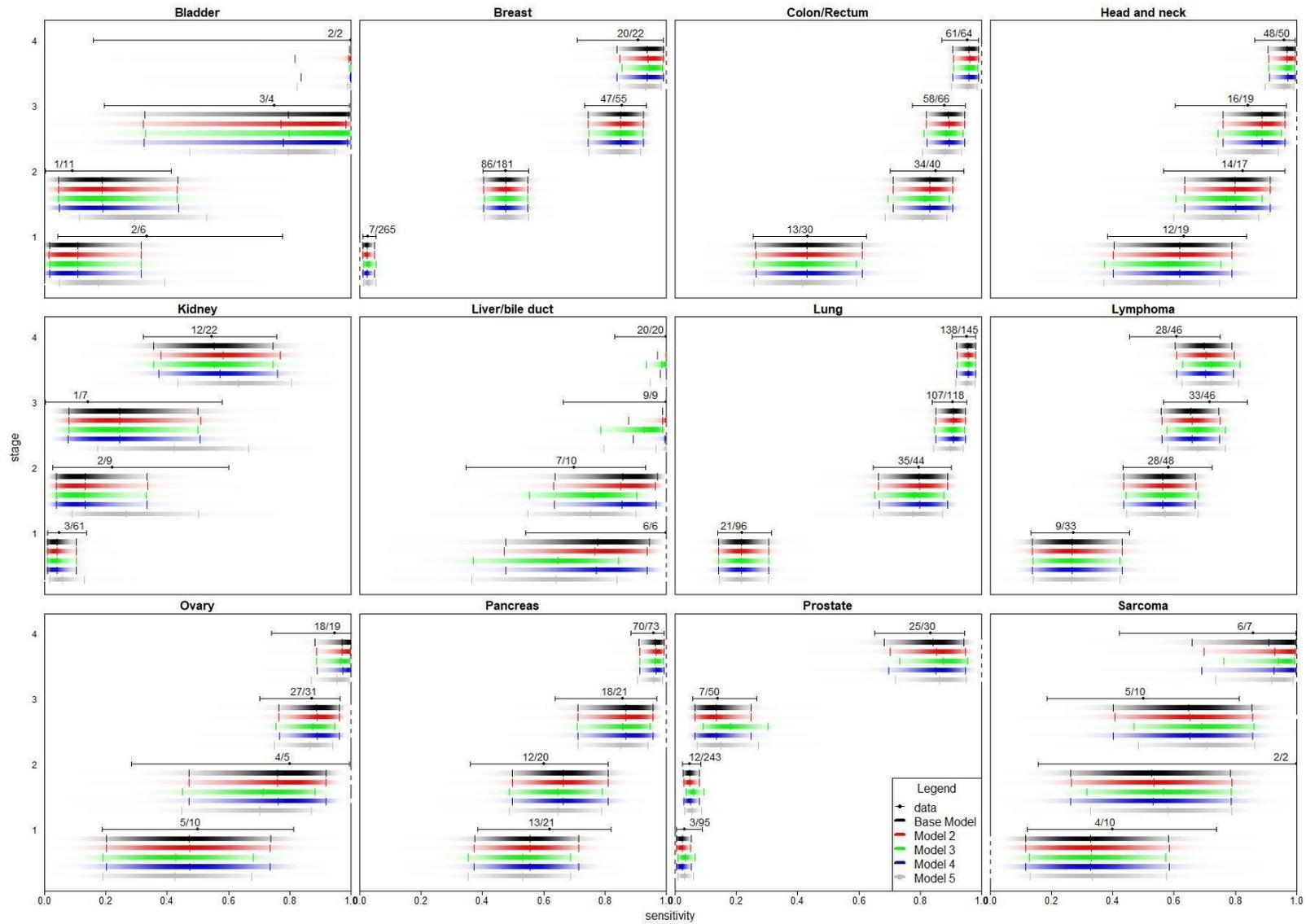

**Figure 2 Data and shrunken estimates of sensitivity for the base model and models 2 to 5 (Part 2)**

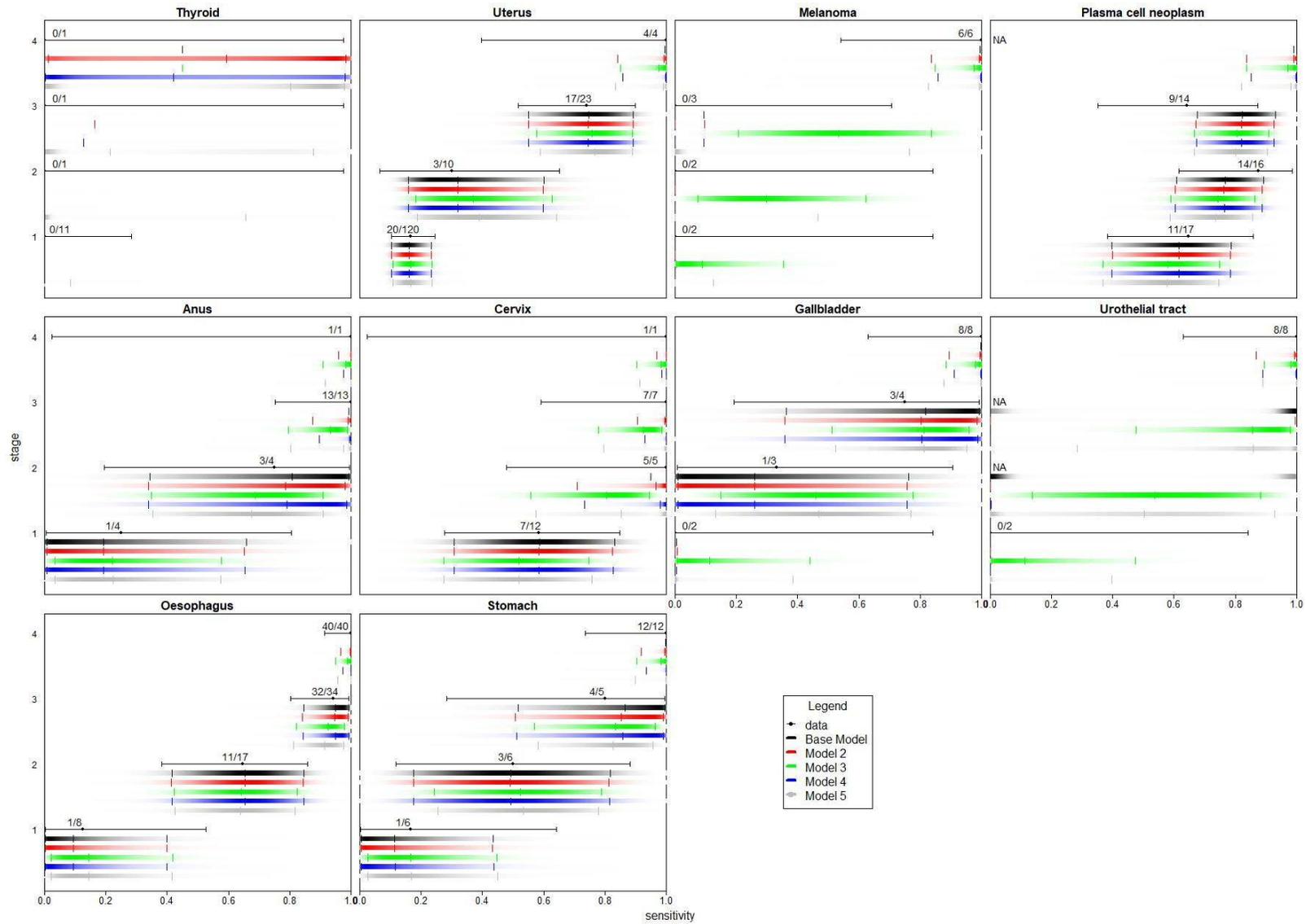

# Supplementary material to
# A Bayesian approach to sharing information on sensitivity of a Multi-Cancer Early Detection test across and within tumour types

**Authors:** Sofia Dias[1], Yiwen Liu[1], Stephen Palmer[2], Marta O Soares[2]

[1] Centre for Reviews and Dissemination, University of York, York, UK

[2] Centre for Health Economics, University of York, York, UK

# Table of contents

# List of Tables

## List of Figures

# S1. ADDITIONAL DATA

*Supplementary material: Dias et al. A Bayesian approach to sharing information on sensitivity of a Multi-Cancer Early Detection test across and within tumour types*

**Table S1. Sensitivity of the Galleri test by cancer type and stage in CCGA3.**

| Code | Cancer type | Overall s/S | Overall % (95% CI) | Stage 1 s/S | Stage 1 % (95% CI) | Stage 2 s/S | Stage 2 % (95% CI) | Stage 3 s/S | Stage 3 % (95% CI) | Stage 4 s/S | Stage 4 % (95% CI) | Missing s/S | Missing % (95% CI) |
|---|---|---|---|---|---|---|---|---|---|---|---|---|---|
| 1 | Bladder | 8/23 | 34.8 (18.8, 55.1) | 2/6 | 33.3 (9.7, 70.0) | 1/11 | 9.1 (1.6, 37.7) | 3/4 | 75.0 (30.1, 95.4) | 2/2 | 100.0 (34.2, 100.0) | 0/0 | NA |
| 2 | Breast | 160/524 | 30.5 (26.7, 34.6) | 7/265 | 2.6 (1.3, 5.4) | 86/181 | 47.5 (40.4, 54.8) | 47/55 | 85.5 (73.8, 92.4) | 20/22 | 90.9 (72.2, 97.5) | 0/1 | 0.0 (0.0, 79.3) |
| 3 | Colon/Rectum | 169/206 | 82.0 (76.2, 86.7) | 13/30 | 43.3 (27.4, 60.8) | 34/40 | 85 (70.9, 92.9) | 58/66 | 87.9 (77.9, 93.7) | 61/64 | 95.3 (87.1, 98.4) | 3/6 | 50.0 (18.8, 81.2) |
| 4 | Head and neck | 90/105 | 85.7 (77.8, 91.1) | 12/19 | 63.2 (41.0, 80.9) | 14/17 | 82.4 (59, 93.8) | 16/19 | 84.2 (62.4, 94.5) | 48/50 | 96 (86.5, 98.9) | 0/0 | NA |
| 5 | Kidney | 18/99 | 18.2 (11.8, 26.9) | 3/61 | 4.9 (1.7, 13.5) | 2/9 | 22.2 (6.3, 54.7) | 1/7 | 14.3 (2.6, 51.3) | 12/22 | 54.5 (34.7, 73.1) | 0/0 | NA |
| 6 | Liver/bile duct | 43/46 | 93.5 (82.5, 97.8) | 6/6 | 100.0 (61.0, 100.0) | 7/10 | 70 (39.7, 89.2) | 9/9 | 100.0 (70.1, 100.0) | 20/20 | 100.0 (83.9, 100.0) | 1/1 | 100.0 (20.7, 100.0) |
| 7 | Lung | 302/404 | 74.8 (70.3, 78.7) | 21/96 | 21.9 (14.8, 31.1) | 35/44 | 79.5 (65.5, 88.8) | 107/118 | 90.7 (84.1, 94.7) | 138/145 | 95.2 (90.4, 97.6) | 1/1 | 100.0 (20.7, 100.0) |
| 8 | Lymphoma | 98/174 | 56.3 (48.9, 63.5) | 9/33 | 27.3 (15.1, 44.2) | 28/48 | 58.3 (44.3, 71.2) | 33/46 | 71.7 (57.5, 82.7) | 28/46 | 60.9 (46.5, 73.6) | 0/1 | 0.0 (0.0, 79.3) |
| 9 | Ovary | 54/65 | 83.1 (72.2, 90.3) | 5/10 | 50.0 (23.7, 76.3) | 4/5 | 80 (37.6, 96.4) | 27/31 | 87.1 (71.1, 94.9) | 18/19 | 94.7 (75.4, 99.1) | 0/0 | NA |
| 10 | Pancreas | 113/135 | 83.7 (76.6, 89.0) | 13/21 | 61.9 (40.9, 79.2) | 12/20 | 60 (38.7, 78.1) | 18/21 | 85.7 (65.4, 95) | 70/73 | 95.9 (88.6, 98.6) | 0/0 | NA |
| 11 | Prostate | 47/420 | 11.2 (8.5, 14.6) | 3/95 | 3.2 (1.1, 8.9) | 12/243 | 4.9 (2.8, 8.4) | 7/50 | 14.0 (7.0, 26.2) | 25/30 | 83.3 (66.4, 92.7) | 0/2 | 0.0 (0.0, 65.8) |
| 12 | Sarcoma | 18/30 | 60.0 (42.3, 75.4) | 4/10 | 40.0 (16.8, 68.7) | 2/2 | 100.0 (34.2, 100.0) | 5/10 | 50.0 (23.7, 76.3) | 6/7 | 85.7 (52.9, 97.8) | 1/1 | 100.0 (20.7, 100.0) |
| 13 | Thyroid | 0/14 | 0.0 (0.0, 21.5) | 0/11 | 0.0 (0.0, 25.9) | 0/1 | 0.0 (0.0, 79.3) | 0/1 | 0.0 (0.0, 79.3) | 0/1 | 0.0 (0.0, 79.3) | 0/0 | NA |
| 14 | Uterus | 44/157 | 28.0 (21.6, 35.5) | 20/120 | 16.7 (11.1, 24.3) | 3/10 | 30.0 (10.8, 60.3) | 17/23 | 73.9 (53.5, 87.5) | 4/4 | 100.0 (51.0, 100) | 0/0 | NA |
| 15 | Lymphoid leukaemia[a] | 21/51 | 41.2 (28.8, 54.8) | | | | | | | | | | |
| 16 | Melanoma | 6/13 | 46.2 (23.2, 70.9) | 0/2 | 0.0 (0.0, 65.8) | 0/2 | 0.0 (0.0, 65.8) | 0/3 | 0.0 (0.0, 56.1) | 6/6 | 100.0 (61.0, 100.0) | 0/0 | NA |
| 17 | Plasma cell neoplasm | 34/47 | 72.3 (58.2, 83.1) | 11/17 | 64.7 (41.3, 82.7) | 14/16 | 87.5 (64.0, 96.5) | 9/14 | 64.3 (38.8, 83.7) | 0/0 | NA | 0/0 | NA |
| 18 | Anus | 18/22 | 81.8 (61.5, 92.7) | 1/4 | 25.0 (4.6, 69.9) | 3/4 | 75 (30.1, 95.4) | 13/13 | 100.0 (77.2, 100.0) | 1/1 | 100.0 (20.7, 100.0) | 0/0 | NA |
| 19 | Cervix | 20/25 | 80.0 (60.9, 91.1) | 7/12 | 58.3 (32.0, 80.7) | 5/5 | 100.0 (56.6, 100.0) | 7/7 | 100.0 (64.6, 100.0) | 1/1 | 100.0 (20.7, 100.0) | 0/0 | NA |
| 20 | Gallbladder | 12/17 | 70.6 (46.9, 86.7) | 0/2 | 0.0 (0.0, 65.8) | 1/3 | 33.3 (6.1, 79.2) | 3/4 | 75.0 (30.1, 95.4) | 8/8 | 100.0 (67.6, 100.0) | 0/0 | NA |
| 21 | Urothelial tract | 8/10 | 80.0 (49.0, 94.3) | 0/2 | 0.0 (0.0, 65.8) | 0/0 | NA | 0/0 | NA | 8/8 | 100.0 (67.6, 100.0) | 0/0 | NA |
| 22 | Myeloid neoplasm[a] | 2/10 | 20.0 (5.7, 51.0) | | | | | | | | | | |
| 23 | Oesophagus | 85/100 | 85.0 (76.7, 90.7) | 1/8 | 12.5 (2.2, 47.1) | 11/17 | 64.7 (2.6, 51.3) | 32/34 | 94.1 (80.9, 98.4) | 40/40 | 100.0 (91.2, 100.0) | 1/1 | 100.0 (20.7, 100.0) |
| 24 | Stomach | 20/30 | 66.7 (48.8, 80.8) | 1/6 | 16.7 (3.0, 56.4) | 3/6 | 50.0 (4.6, 69.9) | 4/5 | 80.0 (37.6, 96.4) | 12/12 | 100.0 (75.8, 100.0) | 0/1 | 0.0 (0.0, 79.3) |
| 25 | Other[b] | 30/59 | 50.9 (39.4, 63.2) | 2/11 | 18.2 (5.1, 47.7) | 3/3 | 100.0 (43.9, 100.0) | 13/18 | 72.2 (49.1, 87.5) | 11/18 | 61.1 (38.6, 79.7) | 1/3 | 33.3 (6.2, 79.2) |
| 26 | Multiple primaries[c] | 16/19 | 84.2 (62.4, 94.5) | 2/2 | 100.0 (34.2, 100.0) | 3/5 | 60.0 (23.1, 88.2) | 6/6 | 100.0 (61.0, 100.0) | 5/6 | 83.3 (43.7, 97.0) | 0/0 | NA |
| 27 | Unknown primary | 17/18 | 94.4 (74.2, 99.0) | 0/0 | NA | 1/1 | 100.0 (20.7, 100.0) | 1/2 | 50.0 (9.5, 90.6) | 13/13 | 100.0 (77.2, 100.0) | 2/2 | 100.0 (34.2, 100.0) |
| | Total | 1453/2823 | 51.5 (49.6, 53.3) | 143/849 | 16.8 (14.5, 19.5) | 284/703 | 40.4 (36.8, 44.1) | 436/566 | 77.0 (73.4, 80.3)) | 557/618 | 90.1 (87.5, 92.2) | 10/20 | 50.0 (29.9, 70.1) |

[a]Not expected to be staged

[b]Includes: adrenal (n=1); ampulla of vater (n=1); brain (n=6); choriocarcinoma (n=1); mesothelioma (n=7); non-melanoma non-BCC/SCC skin cancer (n=2); other/unspecified (n=10); penis (n=1); small intestine (n=13); testis (n=6); thymus (n=2); vagina (n=2); vulva (n=7). Also includes 6 which were not expected to be staged and therefore excluded from data by stage, but still counted in the overall column.

[c]Multiple primaries: If participant had more than 1 cancer at enrolment. If they had multiple primaries and unknown primary, they were counted as multiple primaries to avoid double counting. Highest clinical stage was selected.

**Table S2 Cancer grouping used in class models**

| | Cancer Type | Main Analysis | Sensitivity Analyses | | | |
|---|---|---|---|---|---|---|
| | | Empirically informed | Stage shift[1] | Dwell time[2] | 5-year survival[3] | Cluster informed |
| 1 | Bladder | 1 | 2 | 4 | 2 | 2 |
| 2 | Breast | 2 | 2 | 4 | 3 | 2 |
| 3 | Colon/Rectum | 2 | 1 | 1 | 3 | 1 |
| 4 | Head and neck | 2 | 1 | 2 | 4 | 1 |
| 5 | Kidney | 1 | 2 | 3 | 3 | 3 |
| 6 | Liver/bile duct | 2 | 1 | 3 | 1 | 1 |
| 7 | Lung | 2 | 1 | 1 | 1 | 1 |
| 8 | Lymphoma | 2 | 1 | 2 | 4 | 2 |
| 9 | Ovary | 2 | 1 | 2 | 2 | 1 |
| 10 | Pancreas | 2 | 1 | 3 | 1 | 1 |
| 11 | Prostate | 1 | 3 | 3 | 5 | 3 |
| 12 | Sarcoma | 2 | 2 | 3 | NA | 1 |
| 13 | Thyroid | 1 | 3 | 3 | NA | 4[4] |
| 14 | Uterus | 2 | 2 | 2 | 3 | 2 |
| 15 | Melanoma | 1 | 3 | 4 | 3 | 3 |
| 16 | Plasma cell neoplasm | 2 | NA | 2 | NA | NA |
| 17 | Anus | 2 | 2 | 1 | 3 | 1 |
| 18 | Cervix | 2 | 2 | 2 | 2 | 1 |
| 19 | Gallbladder | 2 | 2 | 3 | 1 | 2 |
| 20 | Urothelial tract | 2 | 3 | 4 | NA | NA |
| 21 | Oesophagus | 2 | 2 | 1 | 1 | 1 |
| 22 | Stomach | 2 | 2 | 3 | 1 | 2 |
| 23 | Other | NA | NA | 4 | NA | NA |
| 24 | Multiple primaries | NA | NA | NA | NA | NA |
| 25 | Unknown primary | NA | NA | NA | NA | NA |

[1] Stage shift groupings based on Table 3 in Sasieni et al.[1] where group 1 show substantial reduction in late stage incidence and group 3 show a small or no shift.
[2] Dwell time groupings based on Supplementary Table S3 in Hubbell et al.[2]
[3] Five-year survival grouping based on Supplementary Figure S1 in Dai et al.[3].
[4] single cancer group – assumed independent from all others for estimation.

## S2. REVIEW OF CLINICAL DETERMINANTS OF HETEROGENEITY

Targeted searches were carried out to identify evidence of clinically relevant determinants of heterogeneity in cancer detection across different cancer types and stages which were used to motivate structural assumptions and assumptions on sharing of information across cancer types and stages in synthesis models of sensitivity by cancer type and stage.

*Supplementary material: Dias et al. A Bayesian approach to sharing information on sensitivity of a Multi-Cancer Early Detection test across and within tumour types*

## 2.1  Methods

A targeted literature review was conducted to identify evidence of clinically relevant determinants of heterogeneity in cancer detection by circulating tumour DNA (ctDNA) blood based multi-cancer early detection (MCED) tests across different cancer types and stages. Evidence on predictors of heterogeneity in ctDNA shedding and the correlation between levels of ctDNA and MCED test performance were summarised narratively. Potential heterogeneity was considered in relation to clinical (cancer type, stage, tumour characteristics) as well as demographic (sex, age) factors.

We searched studies identified in a previous systematic review of the literature on the use of MCED tests in population screening.[4] This was supplemented by an informal search on Google Scholar (using terms such as "ctDNA and MCED test", "factors influencing ctDNA", and "ctDNA and demographics") to identify studies specifically on ctDNA shedding, as the previous review excluded pre-clinical studies reporting no patient-level accuracy data.

## 2.2  Results

Twelve relevant studies were identified, six from a previous systematic review[4] and six from the targeted review of the literature. See Table S3 for details of the studies and the type of evidence available. We identified characteristics that may determine the potential homogeneity or heterogeneity of the sensitivity of the Galleri test across different cancer types and stages to inform plausible information sharing assumptions.

Samples from the CCGA study[5] have been repeatedly used to examine the correlation between ctDNA shedding and test performance. The first and second CCGA sub-studies were used in a model validation study to examine the association between ctDNA shedding and test detection using breast, lung, and colorectal cancers (cancer types preselected based on highest incidence and mortality in US)[6]. All three cancers showed higher likelihood of detectability for cancer cases with higher ctDNA shedding, which also correlated with advancing stage.[6] Similarly, another validation study found a correlation between ctDNA shedding, clinical stage, and detectability.[7] This pattern was clearer for stages 3-4 – with almost all cases of higher ctDNA shedding detected – compared to stages 1-2, where the pattern of detectability was more mixed.[7] In a multivariate analysis examining predictors of cancer signal detection,[8] only ctDNA-related factors predicted test performance (accounting for cancer type and stage) and explained 72% of the variance in cancer signal detection, suggesting that ctDNA shedding may be the main driver of differences in test performance.[8,9]

There is consistent evidence that ctDNA levels are higher in advanced stages of cancer with higher tumour burden compared to localised cancer, regardless of the type of cancer.[6,7,10-15] The size of the tumour has also been associated with the detectability of ctDNA in blood samples: smaller tumours (< 6mm) only contain 0.001% of ctDNA per 10ml of blood, making it unlikely to be detected in blood samples (and also unlikely to progress), compared to larger tumours (10-15mm) which are currently also detectable through imaging (and are more likely to progress).[16]

There is also some evidence to suggest that ctDNA levels measured in blood may be lower in cancers of the central nervous system (e.g., glioma, medulloblastoma) due to the blood-brain barrier.[10, 11, 17, 18] Lower levels of ctDNA have also been found in renal cell, bladder, and kidney cancer,[12, 17, 19] possibly due to the clearance of ctDNA via urine.[20] Endocrine cancer such as thyroid cancer has also been shown to have a lower level of ctDNA.[10-12, 17]

Tumour characteristics and clinical factors have also been associated with levels of ctDNA. Tumour mitotic volume in breast cancer, excessive lesion glycolysis in lung cancer, and larger surface area of deep invading tumours in colorectal cancer, were associated with increased cTF (circulating tumour fractions – a proportion of ctDNA in cell-free DNA [cfDNA]).[6] Tumours with a specific gene mutation (TP53) may also show more ctDNA shedding due to increased metabolic activity or cell turnover.[21] Furthermore, the location of metastases may also be important, as metastatic solid tumours that involve the bone will also shed higher levels of ctDNA compared to tumours that do not spread to the bone.[10]

Demographic differences (e.g., age, sex) were not well reported in the literature; when reported, most investigated associations with survival outcomes rather than variations in ctDNA.[10, 13, 14] One study reported that older adults (≥ 80 years) showed lower ctDNA fraction and quantity compared to younger adults (< 50 years) and that males showed higher ctDNA fraction (13.1%) and quantity (75ng) compared to females (11.3%; 63.2ng); although no age or sex differences were found for cfDNA.[10] Another study found that the fraction of patients with detectable gene alterations increased with age but did not report on how these relate to ctDNA.[14] Other studies which investigated demographic differences in prognosis found mixed results: one found that that older age (> 50 years) and being male were associated with lower rates of survival,[7] whereas another found no differences[13].

**Table S3. Overview of evidence on ctDNA heterogeneity from a targeted review of the literature.**

| | Studies in the literature on ctDNA heterogeneity and correlation with MCED test performance | | | | | |
|---|---|---|---|---|---|---|
| | By cancer type | By stage | By cancer type and stage | Demographic factors | Prognosis | Correlation with test performance |
| Bettegowda 2014[11] | ✓ | ✓ | ✓ (limited) | ✗ | ✓ | ✗ |
| Bredno 2021[6a] | ✗ | ✗ | ✓ (3 cancers only) | ✗ | ✗ | ✓ |
| Bredno 2022[12a] | ✓ | ✓ | ✓ | ✗ | ✓ | ✓ |
| Chen 2021[7a] | ✗ | ✓ | ✗ | ✗ | ✓ | ✓ |
| Huang 2021[10] | ✓ | ✓ | ✗ | ✓ (age, sex) | ✗ | ✗ |
| Jamshidi 2022[8a] | ✗ | ✓ | ✓ (3 cancers only) | ✗ | ✗ | ✓ (multivariate analysis) |
| Liu 2021[9] | ✗ | ✗ | ✗ | ✗ | ✗ | ✓ |
| Patel 2019[13] | ✗ | ✗ | ✓ (pancreatic cancer only) | ✗ | ✓ | ✗ |
| Phallen 2017[14] | ✓ (4 cancers) | ✓ | ✓ (4 cancers with sub-types) | ✓ (age) | ✓ (colorectal cancer) | ✗ |
| Pons-Belda 2021[16a] | ✗ | ✓ (tumour size) | ✗ | ✗ | ✗ | ✗ |
| Venn 2019[15] | ✗ | ✓ | ✓ | ✗ | ✓ | ✓ |
| Zhang 2021[19a] | ✓ | ✓ | ✓ | ✗ | ✓ | ✗ |

[a]Studies identified from the systematic review by Wade et al.[4]

## S3.   SYNTHESIS MODELS

### *3.1   Base model: constraint modelling details*

The base model constrains the sensitivities across cancer types and stages to be monotonically increasing (or equal) with increasing stage for all cancer types *j*

$$\mu_{j1} \leq \mu_{j2} \leq \mu_{j3} \leq \mu_{j4} \tag{1}$$

by specifying

$$\prod_{k=1}^{4} I\left(\mu_{jk} - \mu_{j(k-1)}\right) = 1$$

where *I*(x)=1 if x> 0 and zero otherwise which forces the cancer-type specific (log-odds) sensitivities to be increasing with stage (equation (1)).[22]

### *3.2   Mixture model*

Two mixture models are considered: one where the mixture probabilities depend on cancer type and stage and another where there is a single mixture probability across all stages of the same cancer type. The mixture model based on cancer type only is described as follows: variable *X* determines whether the sensitivity for a particular cancer type *j* is exchangeable with the rest ($X_j = 1$), or not ($X_j = 0$) according to a probability π which is given a Beta prior distribution:

$$X_j \sim \text{Bernoulli}(\pi_j)$$
$$\pi_j \sim \text{Beta}(a_j, b_j)$$

Prior distributions are assigned in a similar way to the cancer/stage mixture model.

### *3.3   Model implementation and fit*

Models were estimated using Markov chain Monte Carlo (MCMC) implemented in JAGS (version 4.3.1) run in R (version 4.4.0) through RStudio (version 2024.04.1+748) using the 'R2jags' package (version 0.8.5). Constraints are implemented using the "ones trick".[22, 23]

Two MCMC chain were run using different starting values. Convergence was checked by inspecting Rhat and by visual assessment of history plots for key model parameters. Models where Rhat for the combined chains is less than or equal to 1.01 (evaluated to 2 decimal places) were considered to have converged.

The residual deviance contributions of each data point were inspected to identify poor fitting points which may indicate conflict between the model predictions and the observed data. The total residual deviance (across all data points) was compared to the number of data points to assess how well the model fits the data.[24, 25] Models with residual deviance close to the total number of data points were preferred, as adequate fit to the data is key to model validity. Cancer types with intermediate sample sizes at each stage are those most likely to have estimates affected by the sharing assumptions (compared to the highest sample cancers) so residual deviance contributions and model predictions for these data points were carefully inspected.

Models with adequate fit (based on the residual deviance) and predictions (based on the shrunken estimates), were compared by looking at differences in Deviance Information Criteria (DIC).[24, 25] The DIC accounts for model complexity as well as fit, with smaller values indicating a more parsimonious model (a model that explains the data well with the smallest number of parameters). Models with similar residual deviance but lower DIC were preferred.

## S4. CLUSTERING OF CANCER TYPES BASED ON TEST PERFORMANCE

K-means clustering (implemented in the R package "stats" version 4.3.1) was applied to CCGA3 data on overall test performance (sensitivity) for each cancer, taking any variations across stages into account. The optimal number of clusters was determined using a combination of the elbow plot and findings from the literature. The elbow plot was used to determine the number of clusters using within-clusters sum of squares (WCSS), which decreases with increasing number of clusters. The optimal number of clusters was selected to be the point at which WCSS decreases slower or in a linear fashion. This was then compared with findings from the literature on ctDNA heterogeneity between different cancer types to ensure that cancers were grouped meaningfully into clusters.

Under the assumption that any similarities or differences between cancer types may likely reflect variations in ctDNA shedding, we clustered cancer types based on their detectability (overall sensitivity), accounting for the increase in detectability with clinical stage (taking sensitivity at each stage into account). Only cancers which were staged (excluded lymphoid leukaemia and myeloid neoplasm) and had available data across all four stages (excluded plasma cell neoplasm and urothelial tract) were included in the analysis. K-means clustering was used to identify meaningful clusters. A 4-cluster solution was deemed to be the most appropriate and consistent with the literature (Table S4).

**Table S4. Grouping of cancers using k-means clustering (4-cluster solution)**

| Cancer | Sensitivity by stage | | | | | Cluster | Description |
|---|---|---|---|---|---|---|---|
| | Overall | 1 | 2 | 3 | 4 | | |
| Colon/Rectum | 82.0% | 43.3% | 85.0% | 87.9% | 95.3% | 1 | High detectability overall and from early to late stage (stage II-IV) |
| Head and neck | 85.7% | 63.2% | 82.4% | 84.2% | 96.0% | | |
| Liver/bile duct | 93.5% | 100.0% | 70.0% | 100.0% | 100.0% | | |
| Lung | 74.8% | 21.9% | 79.5% | 90.7% | 95.2% | | |
| Ovary | 83.1% | 50.0% | 80.0% | 87.1% | 94.7% | | |
| Pancreas | 83.7% | 61.9% | 60.0% | 85.7% | 95.9% | | |
| Sarcoma | 60.0% | 40.0% | 100.0% | 50.0% | 85.7% | | |
| Anus | 81.8% | 25.0% | 75.0% | 100.0% | 100.0% | | |
| Cervix | 80.0% | 58.3% | 100.0% | 100.0% | 100.0% | | |
| Oesophagus | 85.0% | 12.5% | 64.7% | 94.1% | 100.0% | | |
| Bladder | 34.8% | 33.3% | 9.1% | 75.0% | 100.0% | 2 | Medium detectability overall, and high detectability for late stage (stage III-IV) |
| Breast | 30.5% | 2.6% | 47.5% | 85.5% | 90.9% | | |
| Lymphoma | 56.3% | 27.3% | 58.3% | 71.7% | 60.9% | | |
| Uterus | 28.0% | 16.7% | 30.0% | 73.9% | 100.0% | | |
| Gallbladder | 70.6% | 0.0% | 33.3% | 75.0% | 100.0% | | |
| Stomach | 66.7% | 16.7% | 50.0% | 80.0% | 100.0% | | |
| Kidney | 18.2% | 4.9% | 22.2% | 14.3% | 54.5% | 3 | Low detectability overall but high for stage IV only |
| Prostate | 11.2% | 3.2% | 4.9% | 14.0% | 83.3% | | |
| Melanoma | 46.2% | 0.0% | 0.0% | 0.0% | 100.0% | | |
| Thyroid | 0.0% | 0.0% | 0.0% | 0.0% | 0.0% | 4 | Low/no detectability |

Cluster one included colon/rectum, head and neck, liver/bile duct, lung, ovary, pancreas, sarcoma, anus, cervix, and oesophagus cancer, which all showed high detectability overall (60.0%-93.5%), an increase in detectability after stage 1, with high detectability from stage 2 to 4.

Cluster two included bladder, breast, lymphoma, uterus, gallbladder, and stomach cancer, which showed a wider range of detectability overall (28.0%-70.6%), an increase in detectability after stage 2, with high detectability from stage 3 to 4.

Cluster three included kidney, prostate, and melanoma, which showed low detectability overall (11.2%-46.2%), low detectability from stage 1-3, and an increase in detectability at stage 4 only. This is somewhat consistent with the literature, which suggested lower ctDNA shedding in kidney cancer,[12] possibly due to the filtration of ctDNA in the kidney,[17] and higher levels of ctDNA shedding in advanced prostate cancer and melanoma only.[12, 17, 19]

Cluster four only included thyroid cancer, as this was not detected at any stage, and is consistent with the literature which showed lower detectability compared to other types of cancer.[10-12, 17]

## S5. ADDITIONAL RESULTS

Figure S1 shows the observed sensitivities (dot) and 95% CIs (solid lines) for cancer types defined as "Other", "Multiple primaries" and "Unknown" along with the base model predictions (shrunken estimates) presented as density strips with vertical ticks denoting the lower bound of the 95% credible interval (2.5% quantile), the median and the upper bound of the 95% credible interval (97.5% quantile), respectively.

Figure S2 and Figure S3 present the data and model predictions for each cancer type and stage (shrunken estimates) for the base model and models 1A, 1B, 6 and 7. Only cancer types for which sharing assumptions are considered are presented.

**Figure S1 Data and shrunken estimates of sensitivity for the base model for non-specific cancer types**

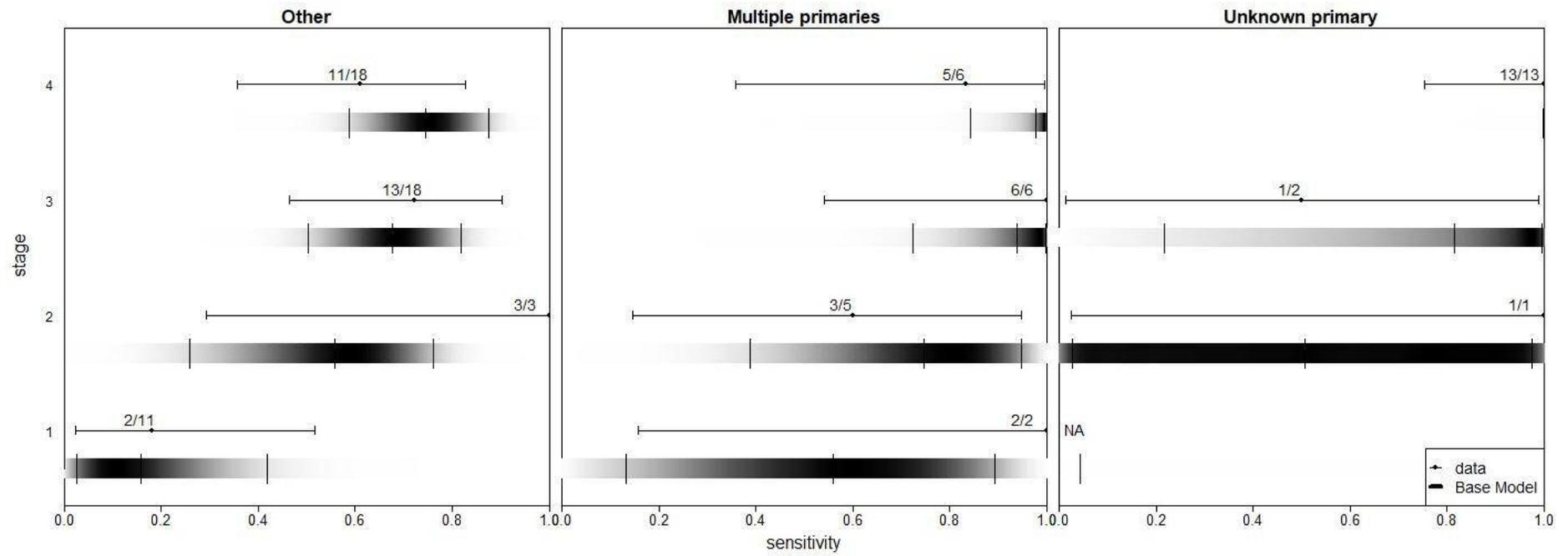

**Figure S2 Shrunken estimates of sensitivity for the base model and models 1A, 1B, 6 and 7 (Part 1)**

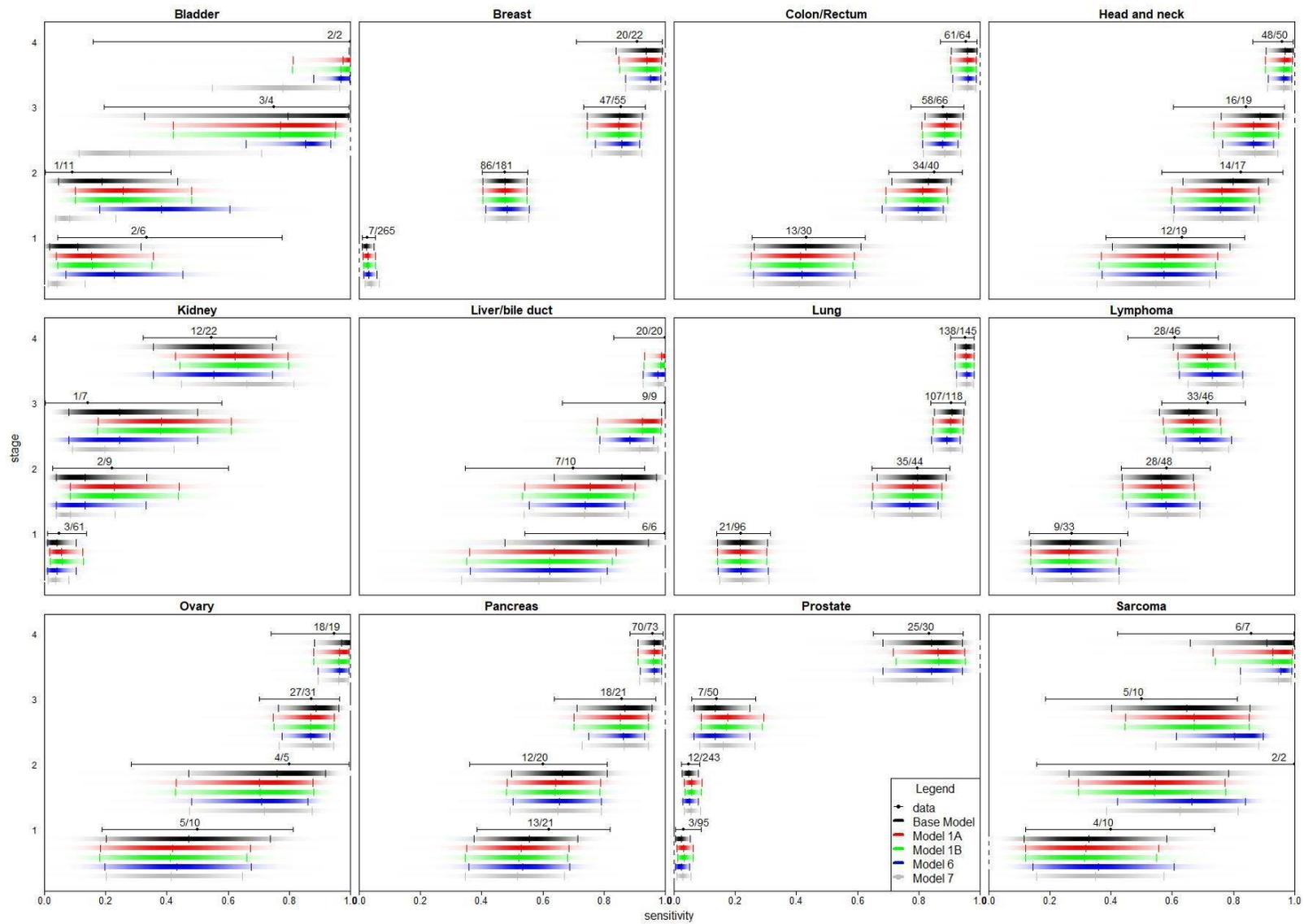

**Figure S3 Shrunken estimates of sensitivity for the base model and models 1A, 1B, 6 and 7 (Part 2)**

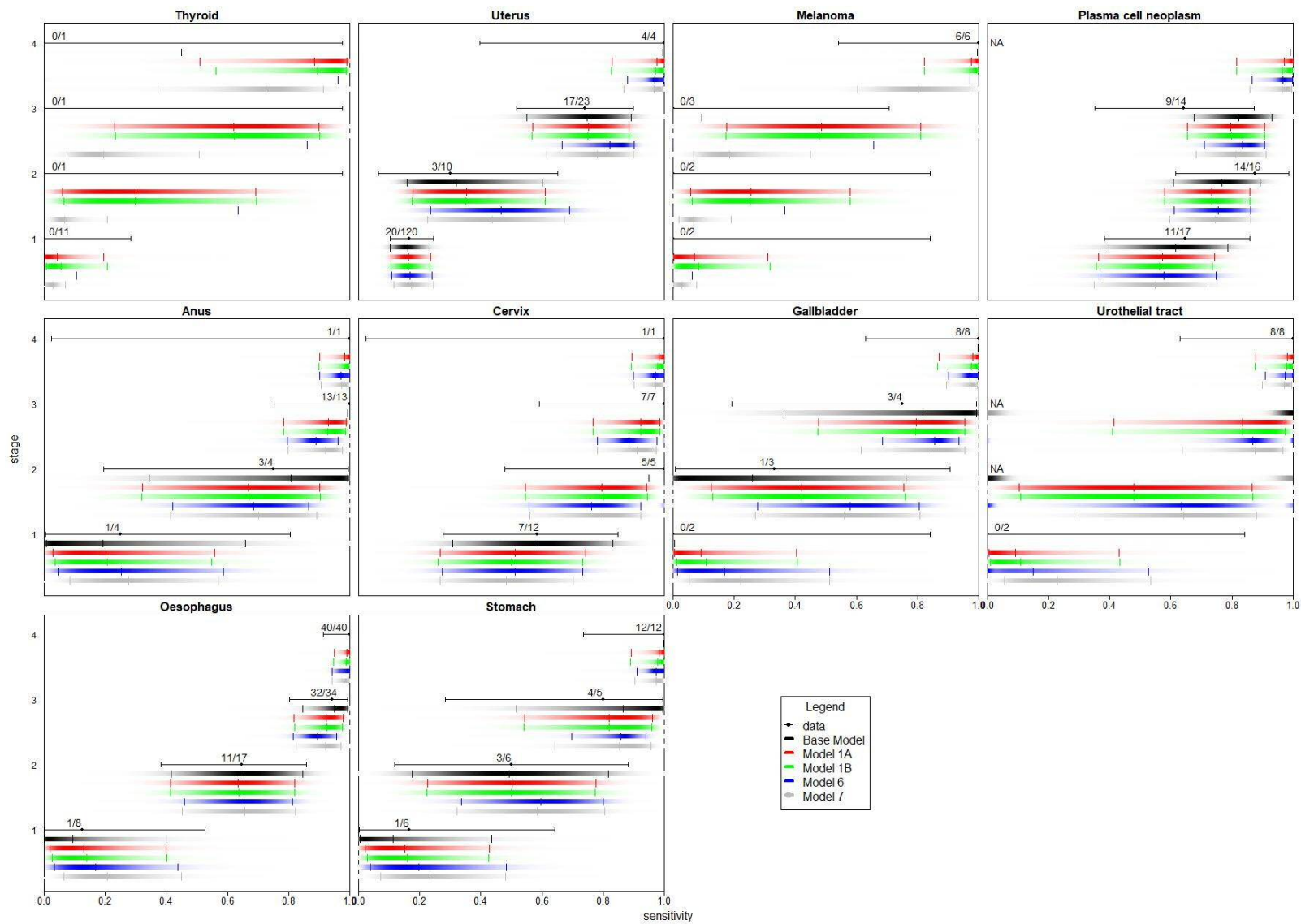

**Table S5 Estimated sensitivity for each cancer type and stage for the base model and values estimated by Sasieni et al.[1]**

| | | stage 1 | | stage 2 | | stage 3 | | stage 4 | |
|---|---|---|---|---|---|---|---|---|---|
| | | mean | [median] | mean | [median] | mean | [median] | mean | [median] |
| cancer type | | (95% CrI) | Sasieni estimate[a] | (95% CrI) | Sasieni estimate[a] | (95% CrI) | Sasieni estimate[a] | (95% CrI) | Sasieni estimate[a] |
| 1 | Bladder | 0.1241 | [0.1087] | 0.2026 | [0.1890] | 0.7562 | [0.7962] | 0.9984 | [1.0000] |
| | | (0.0166, 0.3162) | 0.1765 | (0.0474, 0.4345) | 0.1765 | (0.3280, 0.9917) | 0.7500 | (0.9936, 1.0000) | 1.0000 |
| 2 | Breast | 0.0264 | [0.0252] | 0.4753 | [0.4753] | 0.8462 | [0.8502] | 0.9306 | [0.9362] |
| | | (0.0108, 0.0489) | 0.0264 | (0.4030, 0.5480) | 0.4751 | (0.7449, 0.9241) | 0.8545 | (0.8395, 0.9898) | 0.9091 |
| 3 | Colon/Rectum | 0.4328 | [0.4315] | 0.8252 | [0.8308] | 0.8900 | [0.8927] | 0.9567 | [0.9599] |
| | | (0.2626, 0.6100) | 0.4333 | (0.7116, 0.9061) | 0.8500 | (0.8207, 0.9442) | 0.8788 | (0.9055, 0.9903) | 0.9531 |
| 4 | Head and neck | 0.6116 | [0.6175] | 0.7922 | [0.7989] | 0.8794 | [0.8862] | 0.9640 | [0.9687] |
| | | (0.4036, 0.7889) | 0.6316 | (0.6340, 0.9126) | 0.8235 | (0.7595, 0.9615) | 0.8421 | (0.9068, 0.9952) | 0.9600 |
| 5 | Kidney | 0.0449 | [0.0408] | 0.1467 | [0.1320] | 0.2569 | [0.2446] | 0.5522 | [0.5530] |
| | | (0.0099, 0.1033) | 0.0492 | (0.0390, 0.3333) | 0.1875 | (0.0793, 0.4999) | 0.1875 | (0.3551, 0.7456) | 0.5455 |
| 6 | Liver/bile duct | 0.7586 | [0.7760] | 0.8444 | [0.8590] | 0.9987 | [1.0000] | 1.0000 | [1.0000] |
| | | (0.4767, 0.9441) | 0.8125 | (0.6369, 0.9700) | 0.8125 | (0.9874, 1.0000) | 1.0000 | (1.0000, 1.0000) | 1.0000 |
| 7 | Lung | 0.2188 | [0.2167] | 0.7912 | [0.7967] | 0.9055 | [0.9077] | 0.9539 | [0.9553] |
| | | (0.1420, 0.3068) | 0.2188 | (0.6637, 0.8881) | 0.7955 | (0.8511, 0.9472) | 0.9068 | (0.9188, 0.9805) | 0.9517 |
| 8 | Lymphoma | 0.2720 | [0.2676] | 0.5592 | [0.5623] | 0.6545 | [0.6553] | 0.6975 | [0.6979] |
| | | (0.1378, 0.4305) | 0.2727 | (0.4351, 0.6678) | 0.5833 | (0.5589, 0.7458) | 0.6630 | (0.6036, 0.7897) | 0.6630 |
| 9 | Ovary | 0.4711 | [0.4716] | 0.7416 | [0.7601] | 0.8808 | [0.8872] | 0.9626 | [0.9708] |
| | | (0.2025, 0.7367) | 0.5000 | (0.4712, 0.9172) | 0.8000 | (0.7649, 0.9598) | 0.8710 | (0.8819, 0.9989) | 0.9474 |
| 10 | Pancreas | 0.5526 | [0.5555] | 0.6615 | [0.6641] | 0.8581 | [0.8671] | 0.9611 | [0.9647] |
| | | (0.3742, 0.7139) | 0.6098 | (0.4985, 0.8095) | 0.6098 | (0.7122, 0.9560) | 0.8571 | (0.9103, 0.9916) | 0.9589 |
| 11 | Prostate | 0.0256 | [0.0242] | 0.0513 | [0.0502] | 0.1418 | [0.1365] | 0.8329 | [0.8403] |
| | | (0.0061, 0.0527) | 0.0316 | (0.0286, 0.0806) | 0.0494 | (0.0658, 0.2477) | 0.1400 | (0.6829, 0.9416) | 0.8333 |
| 12 | Sarcoma | 0.3329 | [0.3270] | 0.5267 | [0.5277] | 0.6422 | [0.6476] | 0.8856 | [0.9076] |
| | | (0.1157, 0.5814) | 0.4000 | (0.2624, 0.7848) | 0.5833 | (0.4023, 0.8529) | 0.5833 | (0.6601, 0.9963) | 0.8571 |
| 13 | Thyroid | 0.0000 | [0.0000] | 0.0000 | [0.0000] | 0.0007 | [0.0000] | 0.0308 | [0.0000] |
| | | (0.0000, 0.0000) | 0.0000 | (0.0000, 0.0000) | 0.0000 | (0.0000, 0.0002) | 0.0000 | (0.0000, 0.4499) | 0.0000 |
| 14 | Uterus | 0.1634 | [0.1616] | 0.3362 | [0.3192] | 0.7402 | [0.7469] | 0.9986 | [1.0000] |
| | | (0.1043, 0.2333) | 0.1667 | (0.1588, 0.6001) | 0.3000 | (0.5511, 0.8926) | 0.7391 | (0.9938, 1.0000) | 1.0000 |
| 15 | Melanoma | 0.0000 | [0.0000] | 0.0001 | [0.0000] | 0.0081 | [0.0000] | 0.9986 | [1.0000] |
| | | (0.0000, 0.0000) | 0.0000 | (0.0000, 0.0000) | 0.0000 | (0.0000, 0.0956) | 0.0000 | (0.9948, 1.0000) | 1.0000 |
| 16 | Plasma cell neoplasm | 0.6096 | [0.6165] | 0.7623 | [0.7668] | 0.8174 | [0.8225] | 0.9984 | [1.0000] |
| | | (0.3974, 0.7854) | NA | (0.6077, 0.8916) | NA | (0.6752, 0.9298) | NA | (0.9908, 1.0000) | NA |
| 17 | Anus | 0.2322 | [0.1923] | 0.7663 | [0.8064] | 0.9990 | [1.0000] | 1.0000 | [1.0000] |
| | | (0.0080, 0.6574) | 0.2500 | (0.3442, 0.9918) | 0.7500 | (0.9913, 1.0000) | 1.0000 | (1.0000, 1.0000) | 1.0000 |
| 18 | Cervix | 0.5830 | [0.5876] | 0.9957 | [1.0000] | 1.0000 | [1.0000] | 1.0000 | [1.0000] |
| | | (0.3079, 0.8324) | 0.5833 | (0.9495, 1.0000) | 1.0000 | (1.0000, 1.0000) | 1.0000 | (1.0000, 1.0000) | 1.0000 |
| 19 | Gallbladder | 0.0018 | [0.0000] | 0.2972 | [0.2603] | 0.7777 | [0.8170] | 0.9993 | [1.0000] |
| | | (0.0000, 0.0066) | 0.0000 | (0.0115, 0.7618) | 0.3333 | (0.3646, 0.9922) | 0.7500 | (0.9975, 1.0000) | 1.0000 |
| 20 | Urothelial tract | 0.0011 | [0.0000] | 0.2863 | [0.0000] | 0.7167 | [1.0000] | 0.9997 | [1.0000] |
| | | (0.0000, 0.0000) | 0.0000 | (0.0000, 1.0000) | 0.0000 | (0.0000, 1.0000) | 0.0000 | (1.0000, 1.0000) | 1.0000 |
| 21 | Oesophagus | 0.1242 | [0.0944] | 0.6469 | [0.6529] | 0.9415 | [0.9497] | 0.9998 | [1.0000] |
| | | (0.0037, 0.4003) | 0.1250 | (0.4149, 0.8449) | 0.6471 | (0.8449, 0.9926) | 0.9412 | (0.9994, 1.0000) | 1.0000 |
| 22 | Stomach | 0.1438 | [0.1146] | 0.4944 | [0.4932] | 0.8346 | [0.8650] | 0.9995 | [1.0000] |
| | | (0.0047, 0.4358) | 0.1667 | (0.1755, 0.8175) | 0.5000 | (0.5173, 0.9946) | 0.8000 | (0.9983, 1.0000) | 1.0000 |
| 23 | Other | 0.1747 | [0.1576] | 0.5439 | [0.5572] | 0.6724 | [0.6768] | 0.7418 | [0.7458] |
| | | (0.0248, 0.4175) | 0.1818 | (0.2591, 0.7617) | 0.6923 | (0.5030, 0.8177) | 0.6923 | (0.5876, 0.8743) | 0.6923 |
| 24 | Multiple primaries | 0.5452 | [0.5589] | 0.7253 | [0.7469] | 0.9169 | [0.9378] | 0.9631 | [0.9784] |
| | | (0.1320, 0.8921) | NA | (0.3875, 0.9472) | NA | (0.7237, 0.9975) | NA | (0.8433, 0.9996) | NA |
| 25 | Unknown primary | 0.0079 | [0.0000] | 0.5038 | [0.5058] | 0.7515 | [0.8155] | 0.9995 | [1.0000] |
| | | (0.0000, 0.0424) | NA | (0.0259, 0.9744) | NA | (0.2159, 0.9959) | NA | (0.9984, 1.0000) | NA |

[a] Values from Table S5 in Sasieni et al.[1]; values for lymphoma calculated by the authors due to inconsistent results in Sasieni et al.[1].

**Table S6 Estimated common log-odds, heterogeneity and sensitivity for sharing models 1 to 6.**

|  | Log-odds | | Heterogeneity | | Sensitivity | |
|---|---|---|---|---|---|---|
|  | median | (95% CrI) | median | (95% CrI) | median | (95% CrI) |
| Model 1A | | | | | | |
| Stage 1 | -1.43 | (-2.25, -0.72) | 1.51 | (1.03, 2.37) | 0.19 | (0.10, 0.33) |
| Stage 2 | 0.11 | (-0.51, 0.72) | 1.25 | (0.88, 1.88) | 0.53 | (0.37, 0.67) |
| Stage 3 | 1.36 | (0.74, 2.02) | 1.27 | (0.86, 1.95) | 0.8 | (0.68, 0.88) |
| Stage 4 | 3.25 | (2.44, 4.56) | 1.51 | (0.87, 2.91) | 0.96 | (0.92, 0.99) |
| Model 1B | | | | | | |
| Stage 1 | -1.38 | (-2.03, -0.77) | 1.29 | (1.03, 1.63) | 0.20 | (0.12, 0.32) |
| Stage 2 | 0.11 | (-0.50, 0.72) | | | 0.53 | (0.38, 0.67) |
| Stage 3 | 1.36 | (0.74, 1.99) | | | 0.80 | (0.68, 0.88) |
| Stage 4 | 3.14 | (2.43, 3.95) | | | 0.96 | (0.92, 0.98) |
| Model 2 | | | | | | |
| Stage 4 | 4.02 | (2.69, 6.36) | 2.66 | (1.30, 4.75) | 0.98 | (0.94, 1.00) |
| Model 3 | | | | | | |
| Stage 1 | -1.17 | (-2.01, -0.43) | 1.45 | (0.97, 2.35) | 0.24 | (0.12, 0.39) |
| Stage 2 | 0.36 | (-0.27, 1.00) | 1.20 | (0.82, 1.84) | 0.59 | (0.43, 0.73) |
| Stage 3 | 1.57 | (0.93, 2.26) | 1.21 | (0.80, 1.94) | 0.83 | (0.72, 0.91) |
| Stage 4 | 3.43 | (2.65, 4.82) | 1.28 | (0.68, 2.70) | 0.97 | (0.93, 0.99) |
| Model 4 | | | | | | |
| Stage 4 | 4.97 | (2.94, 7.80) | 3.76 | (1.89, 4.94) | 0.99 | (0.95, 1.00) |
| Model 5 | | | | | | |
| Stage 1 | -1.15 | (-1.96, -0.40) | 1.41 | (0.93, 2.24) | 0.24 | (0.12, 0.40) |
| Stage 2 | 0.35 | (-0.36, 0.93) | 1.05 | (0.52, 1.81) | 0.59 | (0.41, 0.72) |
| Stage 3 | 1.50 | (0.77, 2.10) | 0.95 | (0.38, 1.85) | 0.82 | (0.68, 0.89) |
| Stage 4 | 2.70 | (1.86, 3.85) | 1.20 | (0.43, 2.50) | 0.94 | (0.86, 0.98) |
| Model 6 | | | | | | |
| Stage 1 | -0.89 | (-1.64, -0.22) | 1.22 | (0.80, 2.02) | 0.29 | (0.16, 0.44) |
| Stage 2 | 0.65 | (0.19, 1.10) | 0.70 | (0.40, 1.21) | 0.66 | (0.55, 0.75) |
| Stage 3 | 1.82 | (1.43, 2.23) | 0.47 | (0.04, 0.96) | 0.86 | (0.81, 0.90) |
| Stage 4 | 3.36 | (2.70, 4.57) | 0.89 | (0.04, 2.23) | 0.97 | (0.94, 0.99) |

*Supplementary material: Dias et al. A Bayesian approach to sharing information on sensitivity of a Multi-Cancer Early Detection test across and within tumour types*

**Table S7 Posterior mixture probabilities for Models 5 and 6 (with non-informative priors)**

|   | Cancer type | Model 5 | | | | Model 6 |
|---|---|---|---|---|---|---|
|   |   | Stage 1 | Stage 2 | Stage 3 | Stage 4 |   |
| 1 | Bladder | 0.984 | 0.960 | 0.985 | 0.589 | 1.000 |
| 2 | Breast | 0.910 | 0.987 | 0.988 | 0.985 | 1.000 |
| 3 | Colon/Rectum | 0.981 | 0.976 | 0.986 | 0.984 | 1.000 |
| 4 | Head and neck | 0.972 | 0.982 | 0.986 | 0.983 | 1.000 |
| 5 | Kidney | 0.965 | 0.943 | 0.820 | 0.859 | 0.000 |
| 6 | Liver/bile duct | 0.961 | 0.983 | 0.565 | 0.211 | 0.999 |
| 7 | Lung | 0.984 | 0.981 | 0.986 | 0.985 | 1.000 |
| 8 | Lymphoma | 0.986 | 0.990 | 0.976 | 0.906 | 0.650 |
| 9 | Ovary | 0.981 | 0.985 | 0.986 | 0.979 | 1.000 |
| 10 | Pancreas | 0.978 | 0.989 | 0.987 | 0.982 | 1.000 |
| 11 | Prostate | 0.912 | 0.476 | 0.374 | 0.978 | 0.000 |
| 12 | Sarcoma | 0.983 | 0.988 | 0.984 | 0.981 | 1.000 |
| 13 | Thyroid | 0.084 | 0.311 | 0.504 | 0.722 | 0.079 |
| 14 | Uterus | 0.985 | 0.979 | 0.988 | 0.576 | 1.000 |
| 15 | Melanoma | 0.070 | 0.154 | 0.291 | 0.550 | 0.031 |
| 16 | Plasma cell neoplasm | 0.967 | 0.983 | 0.989 | 0.627 | 1.000 |
| 17 | Anus | 0.980 | 0.982 | 0.535 | 0.276 | 0.999 |
| 18 | Cervix | 0.975 | 0.679 | 0.442 | 0.236 | 0.979 |
| 19 | Gallbladder | 0.487 | 0.980 | 0.985 | 0.508 | 0.999 |
| 20 | Urothelial tract | 0.422 | 0.815 | 0.815 | 0.422 | 0.888 |
| 21 | Oesophagus | 0.977 | 0.987 | 0.971 | 0.269 | 1.000 |
| 22 | Stomach | 0.978 | 0.986 | 0.985 | 0.461 | 1.000 |

Highlighted cells denote mixture probabilities under 50%

**Table S8 Estimated common log-odds, heterogeneity and sensitivity for the class model (classes informed by empirical evidence).**

|   |   | Class | | | |
|---|---|---|---|---|---|
|   |   | 1 | | 2 | |
|   | Stages | median | 95%CrI | median | 95%CrI |
| **Log-odds** | 1 | -3.40 | (-4.57, -2.61) | -0.81 | (-1.31, -0.32) |
|   | 2 | -2.56 | (-3.36, -1.62) | 0.71 | (0.22, 1.20) |
|   | 3 | -1.37 | (-2.20, -0.30) | 1.86 | (1.37, 2.38) |
|   | 4 | 1.15 | (0.33, 2.28) | 3.26 | (2.66, 3.97) |
| **Sensitivity** | 1 | 0.03 | (0.01, 0.07) | 0.31 | (0.21, 0.42) |
|   | 2 | 0.07 | (0.03, 0.17) | 0.67 | (0.56, 0.77) |
|   | 3 | 0.20 | (0.10, 0.43) | 0.87 | (0.80, 0.92) |
|   | 4 | 0.76 | (0.58, 0.91) | 0.96 | (0.93, 0.98) |
| **Heterogeneity**[a] |   | 0.49 | (0.03, 1.40) | 0.84 | (0.64, 1.11) |

[a] within-class heterogeneity on the log-odds scale, common across all stages.
Class definitions in Table S2.

*Supplementary material: Dias et al. A Bayesian approach to sharing information on sensitivity of a Multi-Cancer Early Detection test across and within tumour types*

**Table S9 Model fit statistics for sensitivity analyses**

| Groups based on | Residual deviance[4] | deviance | pD | DIC |
|---|---|---|---|---|
| **Class models** | | | | |
| **Stage shift**[1] | 118.60 | 312.90 | 86.88 | 399.79 |
| **Dwell time**[2] | 112.93 | 307.23 | 82.35 | 389.58 |
| **5-year survival**[3] | 112.46 | 306.76 | 91.56 | 398.32 |
| **Cluster informed groups** | 113.58 | 307.88 | 72.03 | 379.92 |

Class definitions in Table S2.

[1] Stage shift groupings based on Table 3 in Sasieni et al.[1] where group 1 show substantial reduction in late stage incidence and group 3 show a small or no shift.

[2] Dwell time groupings based on Supplementary Table S3 in Hubbell et al.[2]

[3] Five-year survival grouping based on Supplementary Figure S1 in Dai et al.[3]

[4] compare to 100 data points

**Table S10 Estimated common log-odds, heterogeneity and sensitivity for class model with different class definitions (sensitivity analyses).**

| | | Class | | | | | | | |
|---|---|---|---|---|---|---|---|---|---|
| | | 1 | | 2 | | 3 | | 4 | |
| | Stages | median | 95%CrI | median | 95%CrI | median | 95%CrI | median | 95%CrI |
| | | Groups based on: Stage shift[1] | | | | | | | |
| **Log-odds** | 1 | -1.01 | (-1.67, -0.40) | -0.34 | (-1.14, 0.50) | -1.96 | (-3.53, -0.58) | -2.92 | (-4.63, -1.62) |
| | 2 | 1.30 | (0.64, 1.91) | 0.75 | (-0.06, 1.62) | -0.69 | (-2.10, 0.68) | -0.73 | (-2.25, 0.42) |
| | 3 | 2.36 | (1.81, 3.06) | 1.53 | (0.75, 2.43) | 0.56 | (-0.82, 2.00) | 0.85 | (-0.54, 2.10) |
| | 4 | 3.35 | (2.68, 4.35) | 2.62 | (1.69, 3.95) | 2.70 | (1.25, 4.39) | 2.56 | (1.32, 4.66) |
| **Sensitivity** | 1 | 0.27 | (0.16, 0.40) | 0.42 | (0.24, 0.62) | 0.12 | (0.03, 0.36) | 0.05 | (0.01, 0.16) |
| | 2 | 0.79 | (0.65, 0.87) | 0.68 | (0.48, 0.84) | 0.33 | (0.11, 0.66) | 0.32 | (0.10, 0.60) |
| | 3 | 0.91 | (0.86, 0.96) | 0.82 | (0.68, 0.92) | 0.64 | (0.31, 0.88) | 0.70 | (0.37, 0.89) |
| | 4 | 0.97 | (0.94, 0.99) | 0.93 | (0.84, 0.98) | 0.94 | (0.78, 0.99) | 0.93 | (0.79, 0.99) |
| **Heterogeneity**[a] | | 0.29 | (0.02, 0.90) | 0.84 | (0.50, 1.43) | 1.75 | (1.21, 2.67) | 1.01 | (0.46, 2.26) |
| | | Groups based on: Dwell time[2] | | | | | | | |
| **Log-odds** | 1 | -0.21 | (-0.84, 0.43) | -1.91 | (-2.89, -0.99) | -4.22 | (-9.55, -3.11) | | |
| | 2 | 1.05 | (0.43, 1.70) | -0.11 | (-1.00, 0.79) | -3.04 | (-6.09, -2.12) | | |
| | 3 | 1.88 | (1.26, 2.54) | 1.51 | (0.62, 2.50) | -1.88 | (-3.56, -0.42) | | |
| | 4 | 2.80 | (2.15, 3.57) | 3.40 | (2.28, 4.92) | 2.03 | (0.86, 4.25) | | |
| **Sensitivity** | 1 | 0.45 | (0.30, 0.61) | 0.13 | (0.05, 0.27) | 0.01 | (0.00, 0.04) | | |
| | 2 | 0.74 | (0.61, 0.85) | 0.47 | (0.27, 0.69) | 0.05 | (0.00, 0.11) | | |
| | 3 | 0.87 | (0.78, 0.93) | 0.82 | (0.65, 0.92) | 0.13 | (0.03, 0.40) | | |
| | 4 | 0.94 | (0.90, 0.97) | 0.97 | (0.91, 0.99) | 0.88 | (0.70, 0.99) | | |
| **Heterogeneity**[a] | | 0.70 | (0.47, 1.07) | 1.21 | (0.82, 1.83) | 0.34 | (0.01, 3.59) | | |
| | | Groups based on: 5-year survival[3] | | | | | | | |
| **Log-odds** | 1 | -0.70 | (-1.56, 0.01) | -0.46 | (-2.84, 1.70) | -2.30 | (-3.93, -0.88) | -0.29 | (-3.01, 2.46) |
| | 2 | 0.72 | (-0.02, 1.44) | 0.59 | (-1.51, 3.18) | -0.29 | (-1.82, 1.11) | 0.75 | (-1.90, 3.54) |
| | 3 | 2.27 | (1.49, 3.11) | 2.33 | (0.26, 5.30) | 1.06 | (-0.41, 2.51) | 1.27 | (-1.41, 4.05) |
| | 4 | 3.80 | (2.94, 5.07) | 4.82 | (2.22, 10.12) | 2.94 | (1.42, 4.98) | 1.97 | (-0.64, 4.86) |
| **Sensitivity** | 1 | 0.33 | (0.17, 0.50) | 0.39 | (0.06, 0.85) | 0.09 | (0.02, 0.29) | 0.43 | (0.05, 0.92) |

|  |  |  |  |  |  |  |  |  |
|---|---|---|---|---|---|---|---|---|
|  | 2 | 0.67 | (0.50, 0.81) | 0.64 | (0.18, 0.96) | 0.43 | (0.14, 0.75) | 0.68 | (0.13, 0.97) |
|  | 3 | 0.91 | (0.82, 0.96) | 0.91 | (0.56, 1.00) | 0.74 | (0.40, 0.92) | 0.78 | (0.20, 0.98) |
|  | 4 | 0.98 | (0.95, 0.99) | 0.99 | (0.90, 1.00) | 0.95 | (0.81, 0.99) | 0.88 | (0.34, 0.99) |
| **Heterogeneity**[a] |  | 0.61 | (0.24, 1.26) | 1.33 | (0.23, 3.97) | 1.54 | (0.96, 2.68) | 1.41 | (0.57, 3.92) |
| **Cluster informed groups** ||||||||||
| **Log-odds** | 1 | -0.25 | (-0.67, 0.18) | -2.09 | (-3.05, -1.22) | -3.58 | (-5.20, -2.59) |  |  |
|  | 2 | 1.17 | (0.73, 1.60) | -0.39 | (-1.26, 0.45) | -2.73 | (-3.86, -1.59) |  |  |
|  | 3 | 2.15 | (1.73, 2.61) | 1.18 | (0.32, 2.10) | -1.77 | (-2.85, -0.52) |  |  |
|  | 4 | 3.33 | (2.80, 3.98) | 2.51 | (1.53, 3.93) | 1.18 | (0.24, 2.64) |  |  |
| **Sensitivity** | 1 | 0.44 | (0.34, 0.54) | 0.11 | (0.05, 0.23) | 0.03 | (0.01, 0.07) |  |  |
|  | 2 | 0.76 | (0.68, 0.83) | 0.40 | (0.22, 0.61) | 0.06 | (0.02, 0.17) |  |  |
|  | 3 | 0.90 | (0.85, 0.93) | 0.77 | (0.58, 0.89) | 0.15 | (0.05, 0.37) |  |  |
|  | 4 | 0.97 | (0.94, 0.98) | 0.92 | (0.82, 0.98) | 0.77 | (0.56, 0.93) |  |  |
| **Heterogeneity**[a] |  | 0.40 | (0.20, 0.67) | 0.84 | (0.50, 1.43) | 0.48 | (0.02, 1.91) |  |  |

[a] within-class heterogeneity on the log-odds scale, common across all stages.
Class definitions in Table S2.
[1] Stage shift groupings based on Table 3 in Sasieni et al.[1] where group 1 show substantial reduction in late stage incidence and group 3 show a small or no shift.
[2] Dwell time groupings based on Supplementary Table S3 in Hubbell et al.[2]
[3] Five-year survival grouping based on Supplementary Figure S1 in Dai et al.[3].